# Interfacial instability of a planar Interface and diffuseness at the solid-liquid interface for pure and binary materials

Yaw Delali Bensah[1] and J. A. Sekhar[2]

February 10, 2017

## Abstract

Topographical and diffuse interface reconfigurations occur with a change in the solidification rate. In this article we pursue the hypothesis that the interface configuration during solidification is determined by the rate of entropy production in the region between a rigorous solid and rigorous liquid phase. We posit that when an interface begins to migrate, there are several stable configurations that are possible. These include atomistically-planar, diffuse-planar, facet non-planar and cellular non-planar. The configuration and topographical condition that affords the maximum entropy production rate (MEPR) yields the most stable interface configuration. The principle of MEPR is applied to **(1)** describe atomistically smooth and diffuse interfaces, **(2)** provide quantitative results for the *diffuse interface thickness* and the number of pseudo-atomic layers in the interface region, and **(3)** predict the transition from planar to a non-planar facet or non-facet cellular morphology as a function of solidification velocity or temperature gradient.

Numerous experimental investigations spanning over sixty years have failed to *comprehensively* validate any of the existing solid-liquid interface (SLI) growth instability models. With the MEPR model, for the first time, breakdown conditions are predicted with a fair degree of accuracy for a number of binary alloys where no previous theoretical model had predictability. The model considers steady state solidification at close-to and far-from equilibrium conditions.

---

[1] Department of Materials Science and Engineering, University of Ghana, Accra, Ghana.
Email: ydbensah@ug.edu.gh / bensahyad@gmail.com.
[2] University of Cincinnati, Department of Mechanical and Materials Engineering, OH 45221, USA.
MHI Inc. and the Institute of Design and Thermodynamics, Cincinnati, OH 45215, USA.
Email: j.sekhar12@yahoo.com.

**Nomenclature**

*Letter symbols*

$A_f$: area of a solute flux in a liquid ($m^2$)

$A_{SLI}$: area of an interface in a solid-liquid region ($m^2$)

$C_p$: average heat capacity across a solid-liquid interface ($Jm^{-3}K^{-1}$)

$d$: interplanar lattice spacing ($m$)

$dC_{LG}$ or $\Delta C_0$: change in concentration at a solute distance $z$ ($mole\ m^{-3}$)

$D$: Diffusion Coefficient ($m^2s^{-1}$)

$f_s$: fraction of liquid solidified at the solid-liquid interface (dimensionless)

$G_S$: temperature gradient in a solid ($Km^{-1}$)

$G_L$: temperature gradient in a liquid ($Km^{-1}$)

$G_{SLI}$: linear temperature gradient across a diffuse interface ($Km^{-1}$)

$\Delta h_m$: heat of fusion of a solid with defects ($Jm^{-3}$)

$\Delta h_m$: equilibrium heat of fusion ($Jm^{-3}$)

$J_s$: solute flux in a liquid entering a solid-liquid interface ($mole\ s^{-1}$)

$k$: equilibrium partition coefficient obtained from the phase diagram (*dimensionless*)

$k_{eff}$: effective partition coefficient at a solid-liquid interface (*dimensionless*)

$\Delta KE$: gain or loss in kinetic energy ($J$)

$K_L$: thermal conductivity for a rigorous liquid ($Jm^{-1}K^{-1}s^{-1}$)

$K_S$: thermal conductivity for a rigorous solid ($Jm^{-1}K^{-1}s^{-1}$)

$m_L$: slope of the equilibrium liquidus line at the SLI for a binary material ($Km^3mole^{-1}$)

$Q$: lost work potential from the heat generation from a solid-liquid interface ($J$)

$R_g$: molar gas constant ($Jmol^{-1}K^{-1}$)

$S$: Mullins and Sekerka stability constant (*dimensionless*)

$S_f$: flux entropy rate ($JK^{-1}s^{-1}$)

$s_{LG}$: entropy generation density due to solute gradient in a liquid ($Jm^{-3}K^{-1}$)

$s_{SG}$: entropy generation density due to solute gradient in a solid ($Jm^{-3}K^{-1}$)

$\dot{s}_E$: change in entropy generation rate density due to exchange of matter and energy to and from a solid-liquid interface with its surrounding ($Jm^{-3}K^{-1}s^{-1}$)

$\dot{S}_{gen}$: irreversible entropy generation rate in a diffuse region ($JK^{-1}s^{-1}$)

$\dot{S}_{in}$: rate of entropy entering a control volume ($JK^{-1}s^{-1}$)

$\dot{S}_{out}$: rate of entropy leaving a control volume ($JK^{-1}s^{-1}$)

$\dot{s}_{gen}$: total irreversible entropy generated rate density at an interface ($Jm^{-3}K^{-1}$)

$\dot{s}_{LG}$: entropy generation rate density by the solute gradient in a liquid ($Jm^{-3}K^{-1}$)

$(S_{gen})_{max}$: maximum entropy generation due to lost work ($JK^{-1}$)

$dS_{cv}/dt$: total steady state entropy rate in a control volume ($JK^{-1}s^{-1}$)

$ds_{cv}/dt$: total steady state entropy rate density in a control volume ($Jm^{-1}K^{-1}s^{-1}$)

$t$: time ($s$)

$T_{li}$: liquidus temperature at a solid-liquid interface boundary ($K$)

$T_{si}$: solidus temperature at a solid-liquid interface boundary ($K$)

$\Delta T_{SLI}$: temperature difference across a solid-liquid interface ($K$)

$(dC_{LG}/dz)$ or $(\Delta C_0/\delta c)$: change in solute gradient in a liquid ($mole\ m^{-4}$)

$T_m$: melting temperature ($K$)

$T_{av}$: average temperature between $T_{li}$ and $T_{si}$ across a diffuse interface ($K$)

$\Delta T_O$: solidification temperature range (K)

$V$: solidification interface velocity ($ms^{-1}$)
$W_L$: lost work ($J$)
$dz$ or $\delta c$: change in the position length of the solute ($m$)
$Z_{CUT}$: deviation parameter of CUT from experiment at breakdown (*dimensionless*)
$Z_{LST}$: deviation parameter of LST from experiment at breakdown (*dimensionless*)

***Greek symbols***
$\Omega_f$: flux volume ($m^3$)
$\Delta\Omega_S$: volume shrinkage ($m^3$)
$|\Delta\rho_k|$: density shrinkage ($kgm^{-3}$)
$\rho_l$: density of rigorous liquid ($kgm^{-3}$)
$\rho_s$: density of rigorous solid ($kgm^{-3}$)
$\Delta\mu_c$: driving force acting on a solute per melting temperature of solvent medium ($J\ mole^{-1}$)
$\zeta$: solid-liquid interface thickness ($m$)
$\omega_D$: energy of defects ($Jm^{-3}$)
$\Omega_{SLI}$: volume of a solid-liquid interface ($m^3$)
$\dot{\varphi}$: maximum entropy generation rate density for a moving interface ($Jm^{-3}K^{-1}s^{-1}$)
$\eta_G$: driving force diffuseness (*dimensionless*)
$\eta_T$: total diffuseness (*dimensionless*)
$\eta_\alpha$: thermal diffuseness (*dimensionless*)

**Subscripts and acronyms**
CUT: constitutional undercooling theory
LST: linear stability theory
MEPR: maximum entropy production rate
L: liquid
S: solid
$LG$: solute gradients in the liquid
$sg$: solute gradients in the solid
SLI: solid-liquid interface
$HD$: mean heat dissipation at the solid-liquid interface
$f$: facet
$nf$: non-facet
Expt: experiment

## 1. Introduction

The step-kink theory by Frank [1], and Burton, Cabrera, Frank [2] now referred to as BCF theory was the first to describe crystal/liquid interfaces as belonging to one of (a) singular (b) vicinal and (c) non-singular class of surfaces. Cahn and Hilliard [3] later formally analysed the diffuseness of solid-liquid interface for solidification caused by the driving force for a transformation. Earlier studies by Landau [4] and van der Waals [5] had shown that although solid-liquid interfaces could be associated with a thermodynamic potential, a correct equilibrium analysis could only be possible by considering any diffuseness. Cahn [6, 7] also categorized interfaces as belonging to the categories of (a) atomistically smooth or (b) atomistically rough solid-liquid interfaces and further inferred that a transition between smooth and rough could occur with an increase in the overall velocity of transformation. Based on experimental observations it is believed that atomistically smooth interfaces display macroscopic faceting behavior during growth with the appearance of flat sided faces that rely on step-like growth defects for propagation, such as provided by dislocations and ledges. Atomistically rough interfaces on the other hand appear to support continuous growth mechanisms and as a consequence are expected to display topographically smooth but curved interface transitions. However there is no reason that atomistically rough planar interfaces should not transform to macroscopically faceted shapes or vice versa.

When an alloy melt is directionally solidified, a planar morphology is first noted at the solid-liquid interface, usually at a very low velocity of transformation. As the velocity is increased (e.g. by increasing the cooling rate or the Bridgman growth rate) the planar interface becomes unstable to other shapes and transforms to a microscopically diffuse, or a macroscopically jagged/wavy cellular shaped morphology with several variations possible in the topography. When a planar to non-planar topographical transition occurs during solidification (interface growth) it is expected to be a consequence of a thermodynamic driving force and the new shape providing stability compared to other shapes. By careful experimental observations the conditions where the planar to non-planar interface instability becomes noticeable (at optical level magnifications) has been recorded for a vast number of materials and alloys. Additionally, during the growth of crystals from a melt, the onset of diffuseness beyond thermal roughing is often displayed by the solid-liquid interface [3]. The diffuse interface or variations have not been fully factored into the growth topography considerations of a crystal/liquid interface except somewhat in the phase-field literature and previous MEPR discussions [8, 9]. It is instructive to note here that the words roughness and diffuseness appear to have been used interchangeably in the literature when considering a solid-liquid interface structure [3, 10]. In this article, roughness is attributed to thermal influences whereas full diffuseness is attributed to disordering by both thermal and other driving forces for interface migration.

An interface roughness criterion/model developed by Jackson [10] compares *the* bond enthalpy to the temperature (thermal) roughening, $K_B T_m$ at the melting point $T_m$, where $K_B$ is the Boltzmann constant. This model suggests that when the roughness criterion is greater than 2 then an atomistically sharp interface is predicted i.e. smooth macroscopic features are expected, and when the roughness is less than 2 then an atomistically rough interface is expected. Although this model has had some success there are notable problems, the most significant one being for succinonitrile which is predicted by this model to be faceted but has not shown any such tendencies. The extent of thermal roughening is labelled $\eta_\alpha$ in this article. The $\eta_\alpha$ is the inverse of the Jackson criterion number and also corresponds to the number of interface atomic layers between the rigorous solid or rigorous liquid regions.

Cahn et al [6, 7] have shown that interface diffuseness (beyond thermal roughening) can also be enabled by an increased driving force for the transformation (i.e. an increased solidification velocity). In this article, this type of roughening is referred to as driving force diffuseness $\eta_G$ (where $\eta_G$ is the number of pseudo-atomic-planes of 'roughness' caused by the free energy difference required to drive the interface). The total diffuseness $\eta_T = (\eta_\alpha + \eta_G)$ is defined as the sum of the diffuse layer roughness from both the driving force and thermal energy.

As mentioned above, the region between the solidus and liquidus boundaries in an alloy, during solidification may additionally contain macroscopically identifiable variations in topography in addition to roughening. The appearance of a cellular or jagged morphology from a planar interface, especially for binary-alloy materials is traditionally at least thought to depend on the material composition, $C_O$ (wt% or mole/m$^3$), velocity $V$ (m/s) of the growing interface and the temperature gradient $G_L$ (*K/m*) in the liquid. Also the process conditions that lead to distinct interface transitions are the interface velocity, temperature gradient, composition. These variables are commonly subscripted with the symbol $_{(C)}$ or $_{(B)}$ [11-19] to indicate a transition. In this article the subscript $_{(C)}$ is used to denote the critical condition. Although a number of theoretical models have been proposed to explain and predict the critical condition for the interface breakdown, interface roughening is not normally considered as a variable in these models except at very high rates of solidification. The two most widely employed models that describe the interface instability from planar to non-planar are the constitutional undercooling (CUT) [20] and linear stability theory based model (LST) [21].

The CUT model was proposed qualitatively by Rutter and Chalmers [22] and later quantitatively described by Tiller, Rutter, Jackson, and Chalmers [20]. This model describes the interface instability (from planar to non-planar) as being triggered by a region of constitutionally undercooled liquid that forms ahead of the solid-liquid interface during growth because of solute partitioning. For a binary alloy the CUT criterion for instability is given as:

$$\left(\frac{V}{G_L}\right)_C = \frac{D_L}{\Delta T_O} \qquad (1)$$

where $G_L$ (K m$^{-1}$) is the temperature gradient in the liquid, $D_L$ (m$^2$ s$^{-1}$) is the solute diffusion coefficient in the liquid and $\Delta T_O$ (K) is the equilibrium solidification range (T$_l$-T$_S$) for a liquid at composition $C_O$ (mole m$^{-3}$). Also $T_l$ (K) and $T_S$ (K) are the equilibrium liquidus and solidus temperatures shown in the equilibrium phase diagrams. The ratio of experimentally measured critical (V/G$_L$)$_{exp}$ to (D$_L$/ΔT$_O$), for the CUT criterion is one (equation 1). Thus if correct, the model may be used to infer the diffusion constant. However, it has been recently noted [23] that very significant deviations are noted in the predicted diffusion constants made by the CUT theory. For this article, the numerical deviation from experimentally measured breakdown is labelled as the CUT deviation parameter $Z_{CUT}$ (dimensionless) - shown in table 1 for several binary alloy systems.

In 1964 Mullins and Sekerka [21] proposed the linear stability theory (LST) which considered the stability of a planar interface to a perturbation of infinitesimal amplitude. The interface is unstable if any wavelength of a sinusoidal perturbation grows, and alternately is stable if none of the perturbations are able to grow. This LST criterion gives the instability criterion for a binary material as:

$$\left(\frac{V}{G_L}\right)_C = \frac{D_L}{\Delta T_O}\frac{2\,K_L}{(K_S+K_L)\,S} \tag{2}$$

where S (no units) is Mullins and Sekerka stability constant [21] which is equal to one for low velocities, $K_L$ and $K_S$ (J m$^{-1}$K$^{-1}$s$^{-1}$) are the thermal conductivities for the rigorous solid and liquid respectively.

Bensah et al. [23] and De Cheveigne et al. [15] have shown that there is also a significant deviation that is noted when comparing the LST model predictions with experiments. The numerical LST deviation from experimentally measured breakdown is labelled as the $Z_{LST}$ (dimensionless) is also shown in table 1. A study by Burgeon et al. [24] on in-situ microgravity interface imaging during the ordering of a cellular array structure, has concluded that the cause of interface dynamics and breakdown are more than just on account of the undercooled liquid ahead of the interface. A recent experimental study by Inatomi et al. [25] has further cast doubt on whether an undercooled liquid or solute pile-up ahead of the interface is always present. They have argued persuasively that none of the theories for breakdown [20, 21] may be correct. For an interface topographical instability in the case of facet prone materials, a strain accumulation model [26] has also been considered as describing the interface breakdown. However, Inatomi et al. [25] argue also against a general strain model as the cause for the instability. For the conditions where the interface breakdown occurs at high velocities especially for very low alloy composition materials or with very low temperature gradients (see tables 1 and 2), both the CUT and LST models lose even more predictive capability [23]. Additionally, it should be noted the CUT and LST models do not address the facet/non-facet diffuseness at a molecular level although clearly this is an important feature of instability albeit for a smooth to rough interface but mostly only observable along with a topographical instability from planar to non-planar.

The analyses of solid-liquid interfaces by Sekhar [9], Hill [27], Kirkaldy [28], and Martyushev et al. [29] have shown that the interface instability may be analysed with the maximum entropy production *rate* (MEPR) postulate. The theoretical foundation of MEPR was first given by Ziman [30] and Ziegler [31, 32]. Such a formulation is widely believed by many as an extension of the second law of thermodynamics and also regarded by some as a possible new thermodynamic law by itself that reveals pathway selection rules for a dynamic system [9, 33-37]. Whereas a minimization of the rate of entropy production is required for equilibrium conditions in a closed system, Sekhar [9] has pointed out that the *maximization* of the rate of entropy production within an open control-volume is required for the description of systems that continuously interact with the surroundings. The most stable diffuseness or topographical features are related to such maximization.

This article describes a new solidification model based on the maximum entropy generation rate principle which considers the lost work potential as the criterion for the stability of any interface configuration at the solid-liquid interface. The lost work potential is a consequence of free energy dissipation process that is required for the phase change. In the earliest MEPR formulation [9], the calculation of the interface temperature difference between a rigorous solid and rigorous liquid was possible only for a few conditions. It is shown below that an extended MEPR model is able to quantitatively relate interface thickness to the diffuseness for binary alloys. The model is also able to unify the driving force diffuseness and the thermal diffuseness (into a total diffuseness number) into one expression which can quantitatively guide stability considerations based on features that describe the highest entropy rate production at the interface. It should be noted that the MEPR analysis that is described below is only rigorously valid at a steady state conditions.

The model predictions are tested with experimental data available from numerous published studies. The driving force diffuseness and thermal diffuseness unification enables the model to also be predictive of the velocity and temperature gradient dependency that have been noted for facet/non-facet transition (*f/nf*) in many solidification studies. A considerable number of topographical transitions in dilute binary materials are compared with an MEPR instability criterion that fully provides the sufficient condition for interface instability from planar to non-planar by considering the interface diffuseness parameters.

## 2. MEPR model

### 2.1. Entropy generation at the solid-liquid interface

Consider the changeover region between a solidifying liquid to solid in directional solidification (DS) system that has a finite dimension over which a temperature gradient and other gradients are established. This changeover zone is called a solid-liquid interface (SLI) region with a thickness $\zeta$ (m). The heat of fusion of the solid with defects, $\Delta h_m$ (J m$^{-3}$) and the equilibrium heat of fusion $\Delta h_{sl}$ (J m$^{-3}$), within the SLI are related by [9]:

$$\Delta h_{sl} \; = \; \Delta h_m \; + \; \omega_D \qquad (3a)$$

where $\omega_D$ (J m$^{-3}$) is the energy of defects (such as grain boundaries or dislocations) per unit volume. For this article, it is assumed that $\omega_D$ is a relatively small term - equation (3a) becomes:

$$\Delta h_{sl} \; = \; \Delta h_m \qquad (3b)$$

Note that by assuming that $\omega_D$ is small does not imply that the lost work potential (discussed further below) is small. The interface region is bound by rigorous solid and rigorous liquid phases on either side [9]. The entropy rate balance for the control volume is given by [9]:

$$\frac{dS_{cv}}{dt} \; = \; \dot{S}_{in} - \dot{S}_{out} \; + \; \dot{S}_{gen} \qquad (4)$$

where $\frac{dS_{cv}}{dt}$ (J K$^{-1}$s$^{-1}$) is the total steady state entropy rate change in the control volume, $\dot{S}_{in}$ (J K$^{-1}$s$^{-1}$) and $\dot{S}_{out}$ (J K$^{-1}$s$^{-1}$) are the rate of entropy entering and leaving the control volume respectively, and $\dot{S}_{gen}$ (J K$^{-1}$s$^{-1}$) is the *irreversible* entropy generation rate in the diffuse region. The rates of entropy entering ($\dot{S}_{in}$) and leaving ($\dot{S}_{out}$) the control volumes are given by:

$$\dot{S}_{in} \; = \; A_{SLI} \, V \left( \frac{\Delta h_{sl}}{T_{li}} + s_{LG} \; + \; s_{SG} \right) \qquad (5)$$

$$\dot{S}_{out} \; = \; A_{SLI} \, V \left( \frac{\Delta h_m}{T_{si}} + s_{SG} \right) \qquad (6)$$

where the subscripts $(_{LG})$ and $(_{SG})$ refer to solute gradients in the liquid and solid respectively, V (m s$^{-1}$) is the solidification interface velocity, $s_{LG}$ (J m$^{-3}$K$^{-1}$) is the entropy generation density due to solute gradient in the liquid, $s_{SG}$ (J m$^{-3}$K$^{-1}$) is the entropy generation density due to solute gradient in the solid, A$_{SLI}$ (m$^2$) is the area of the interface in the solid-liquid region and, T$_{li}$ (K) and T$_{si}$ (K) are liquidus and solidus temperatures at the SLI boundaries respectively. It is also assumed that the thermal gradient (similar to assumptions made in the LST model) across the solid-liquid interface is linear and expressed as:

$$\Delta T_{SLI} = T_{li} - T_{si} = \zeta\ G_{SLI} \tag{7}$$

where G$_{SLI}$ (K m$^{-1}$) is the linear temperature gradient across the diffuse interface, and $\Delta$T$_{SLI}$ (K) is the temperature difference across the SLI. The volume of the solid-liquid interface $\Omega_{SLI}$ (m$^3$) is given as:

$$\Omega_{SLI} = A_{SLI}\ \zeta \tag{8}$$

By combining equations (5), (6) and (7) into equation (4) yields the control volume expression at steady state as:

$$\frac{dS_{cv}}{dt} = \left(\frac{A_{SLI}\ V\Delta h_{sl}}{T_{li}} + A_{SLI}Vs_{LG} + A_{SLI}Vs_{SG}\right) - \left(\frac{A_{SLI}\ V\Delta h_m}{T_{si}} + A_{SLI}Vs_{SG}\right) + \dot{S}_{gen} \tag{9a}$$

Further rearranging equation (9a) gives:

$$\frac{dS_{cv}}{dt} = \frac{A_{SLI}\ V\Delta h_{sl}}{T_{li}} - \frac{A_{SLI}\ V\Delta h_m}{T_{si}} + \Omega_{SLI}\ \dot{s}_{LG} + \dot{S}_{gen} \tag{9b}$$

where $\dot{s}_{LG}$ (J m$^{-3}$K$^{-1}$) is the entropy generation rate density by the solute gradient in the liquid. If equation (9b) is divided by the volume of the solid-liquid interface as expressed in equation (8) one obtains,

$$\frac{ds_{cv}}{dt} = \frac{V\Delta h_{sl}}{\zeta\ T_{li}} - \frac{V\Delta h_m}{\zeta\ T_{si}} + \dot{s}_{LG} + \dot{s}_{gen} \tag{10}$$

where $\dot{s}_{gen}$ (J m$^{-3}$K$^{-1}$) is the total entropy generation rate density at the interface and $\frac{ds_{Cv}}{dt}$ (J m$^{-3}$K$^{-1}$) becomes the total steady state entropy rate density in the control volume. Substituting of equation (3b) into equation (10) and applying the steady state condition, $\frac{ds_{Cv}}{dt} = 0$, then the total entropy generation rate density at the interface (in the SLI region) becomes:

$$\dot{s}_{gen} = \left( \frac{V \Delta h_{sl}}{\zeta\, T_{si}} - \frac{V \Delta h_{sl}}{\zeta\, T_{li}} \right) - \dot{s}_{LG} \qquad (11)$$

The expression in parenthesis in equation (11) is the entropy generation rate density $\dot{s}_E$ (J m$^{-3}$K$^{-1}$) which describes the new entropy generated due to exchange of matter, and bond formation [9] which in its simplified form may be written as:

$$\dot{s}_E = \frac{V\, \Delta h_{sl}\, G_{SLI}}{T_{li} \cdot T_{si}} \qquad (12)$$

## 2.2. Entropy generation from the solute gradient in the liquid

For steady state conditions, the solute flux $J_s$ (mole s$^{-1}$) in the liquid entering the interface for a given flux area $A_f$ (m$^2$) is related to the Fick's first law of diffusion [38] as:

$$J_s = -A_f\, D_L\, \left( \frac{dC_{LG}}{dz} \right) \qquad (13)$$

where $\left( \frac{dC_{LG}}{dz} \right)$ (mole m$^{-4}$) is the change in solute gradient in the liquid, dz (m) is the change in the position length of the solute, and $dC_{LG}$ (mole m$^{-3}$) is the change in concentration at a distance, $z$ from the interface. The solute gradient in the liquid can be replaced with ($-\Delta C_O/\delta c$) [39] where $\delta_c$ (m) is the diffusion boundary layer and the negative sign represents the depletion of solute along the distance, $z$. Entropy is also generated when the solute in the liquid travels across the interface to form a solid through an established solute gradient. The driving force $\Delta\mu_c$ (J mole$^{-1}$) associated with the solute gradient is given as [9]:

$$\Delta\mu_C = R_g\, T_m\, ln(1/k) \qquad (14)$$

where $R_g$ (J mole$^{-1}$ K$^{-1}$) is the molar gas constant, $T_m$ (K) is the melting temperature and $k$ (dimensionless) is the equilibrium partition coefficient obtained from the phase diagram (in concentration units of mole m$^{-3}$). Although k is non-dimensional the numerical value depends on the concentration units chosen. However, for the entropy generation calculations this is multiplied by the composition difference. It is also recognized that when comparing interface configurations for stability the value of k for a diffuse interface based configuration will be different than that when the interface has a atomistically smooth topography. Multiplying equation (13) by equation (14) and diving by the melting temperature T$_m$ (K) of the material gives the flux entropy rate S$_f$ (J K$^{-1}$s$^{-1}$) as:

$$S_f = A_f \ D_L R_g \ \left(\frac{\Delta C_O}{\delta_C}\right) ln(1/k) \qquad (15)$$

The change in solute gradient in the liquid $\Delta C_O$ (mole m$^{-3}$), the flux volume $\Omega_f$ (m$^3$) and the diffusion boundary layer $\delta_C$ (m) are respectively given as:

$$\Delta C_O = \frac{\Delta T_O}{m_L} \qquad (16)$$

$$\Omega_f = A_f \ \delta_C \qquad (17)$$

$$\delta_C = \frac{2 \ D_L}{V} \qquad (18)$$

where m$_L$ (Km$^3$ mole$^{-1}$) is the slope of the equilibrium liquidus line at the solid-liquid boundary for a binary material obtained from the phase diagram. Now rearranging equations (16) and (18) into equation (15) and dividing by equation (17) gives the entropy rate density which describes the force-flux entropy generated by the existence (support) of maintaining the solute gradient as [9]:

$$\dot{s}_{LG} = \frac{\Delta T_O}{D_L} \frac{V^2 R_g}{4} \frac{ln(1/k)}{m_L} \qquad (19)$$

For the entropy generation inside the boundaries of the solid liquid zone this gradient entropy reduces the total amount of the irreversible entropy generated as may be noted from equation (4).

### 2.3. Entropy generation and the conversion of kinetic energy

The overall transformation includes a density change $|\Delta\rho_k|$ (kg m$^{-3}$) given by:

$$|\Delta\rho_k| = \left|\frac{\rho_l\,\Delta\rho}{\rho_s}\right| \qquad (20)$$

where $\Delta\rho_k$ (kg m$^{-3}$) is the overall density shrinkage expressed as $\Delta\rho_k = \rho_l\,\Delta\rho\,/\rho_s$, and $\Delta\rho$ (kg m$^{-3}$) is the density change from liquid to solid ($\rho_s$-$\rho_l$); $\rho_s$ (kg m$^{-3}$) and $\rho_l$ (kg m$^{-3}$) are the densities of rigorous solid and liquid respectively. For the rest of this derivation the modulus sign for the density shrinkage is omitted. The volume shrinkage $\Delta\Omega_S$ (m$^3$) associated with the transformation is given as:

$$\Delta\Omega_S = A_{SLI}\ \zeta\ \Delta\rho_k \qquad (21)$$

The change in kinetic energy of a moving liquid transforming into solid is:

$$\Delta KE = \frac{1}{2}\,\rho_l\,\Delta\Omega_S\,V^2 \qquad (22)$$

Placing equations (20) and (21) into equation (22) gives the overall gain or loss in kinetic energy $\Delta KE$ (J) of the transforming liquid entering into the SLI as:

$$\Delta KE = \frac{A_{SLI}\,\zeta\ \Delta\rho_k\,V^2}{2} \qquad (23)$$

The moving interface dissipates free energy equal to the lost work, $W_L$ (J) as given in equation (24). The lost work is equivalent to the loss in kinetic energy given in equation (25), which is obtained by combining equations (23) and (24). *The key hypothesis in this article is that MEPR is operative with maximum entropy generation rate density, $\dot{\varphi}_{max}$ (J m$^{-3}$K$^{-1}$s$^{-1}$) within the SLI, which is then predictive of the most stable morphology.*

$$W_L = T_{av}\big(S_{gen}\big)_{max} \qquad (24)$$

$$\big(S_{gen}\big)_{max} = \frac{A_{SLI}\,\zeta\ \Delta\rho_k\,V^2}{2\,T_{av}} \qquad (25)$$

where $(S_{gen})_{max}$ (J K$^{-1}$) is the maximum entropy generation due to the lost work and $T_{av}$ (K) is the average temperature between $T_{li}$ and $T_{si}$ across the diffuse interface. Following the work term introduced in equation (24) and reference [9], the main assumption in this article is that the gain in kinetic energy is converted to heat which is further converted to some work subject now to the limitation of the second law of thermodynamics. The lost work potential from the heat generation, Q (J) is:

$$Q = A_{SLI} \, \zeta \, C_p \, \Delta T_{SLI} \qquad (26)$$

where $C_p$ (J m$^{-3}$K$^{-1}$) is the average heat capacity across the SLI ($\zeta$). With equation (26), the equivalent entropy generation through heat dissipation, $(S_{gen})_{HD}$ (J K$^{-1}$) may be approximated as:

$$\left(S_{gen}\right)_{HD} = A_{SLI} \, \zeta \, C_p \, \frac{\Delta T_{SLI}}{T_{av}} \qquad (27)$$

where the subscript (HD) indicates the heat dissipation. The temperature gradient at the SLI ($G_{SLI}$) maybe approximated as:

$$G_{SLI} = \frac{(G_S + G_L)}{2} \qquad (28)$$

where $G_S$ (K m$^{-1}$) and $G_L$ (K m$^{-1}$) are the temperature gradients in the solid and liquid respectively. The maximum entropy generation due to the lost work is equal to the equivalent entropy generation through heat dissipation. Combining equations (25) and (27), and substituting in equations (7) and (28) gives the heat capacity:

$$C_p = \frac{\Delta \rho_k \, V^2}{2 \, \zeta \, G_{SLI}} \qquad (29)$$

The maximum entropy generation rate density (MEPR) (no solute partitioning case), $\dot{\varphi}_{max}$ (J m$^{-3}$K$^{-1}$s$^{-1}$) (eqn 31), is now obtained by multiplying equation (29) by the change in the fraction of the liquid solidified per second (equation 30).

$$\frac{df_s}{dt} = \frac{V}{\zeta} \tag{30}$$

$$\left(C_p \frac{df_s}{dt}\right)_{max} = \frac{d\varphi_{max}}{dt} = \dot{\varphi}_{max} \tag{31a}$$

$$\dot{\varphi}_{max} = \frac{\Delta\rho_k}{2} \frac{V^3}{\zeta^2 G_{SLI}} \tag{31b}$$

Where $f_s$ (dimensionless) is the fraction solidified and t (s) is time. Thus $\dot{\varphi}_{max}$ becomes a function of $\zeta$, V and $G_{SLI}$. When partitioning is feasible, the maximum entropy generated rate density can be expressed by combining with equations (12) and (19) into equation (11) as:

$$\dot{\varphi}_{max} = \frac{V \Delta h_{sl} G_{SLI}}{T_{li} \cdot T_{si}} - \frac{\Delta T_O}{D_L} \frac{V^2 R_g \ln(1/k)}{4 \ m_L} \tag{32}$$

The maximization of the entropy generation rate equation (32) is the pathway or interface selection that the interface will prefer. From equation (32), it is noted that $\dot{\varphi}_{max}$ is a function of $\zeta$, V, $G_{SLI}$, $D_L$ and k.

### 2.4. Interface thickness, diffuseness and stability of an atomistically smooth interface for pure materials.

Pure materials may grow in an atomistically smooth, diffuse or smooth but jagged manner. Reference is made to equation (32) where the last term is set to zero for pure materials. From this, the diffuse interface thickness $\zeta$ is given as:

$$\zeta = \frac{V}{G_{SLI}} \left(\frac{\Delta\rho_k}{2} \frac{T_m^2}{\Delta h_{sl}}\right)^{\frac{1}{2}} \tag{33a}$$

$$\zeta = \frac{V}{G_{SLI}} \frac{1}{\sqrt{M}} \tag{33b}$$

The expression in the parenthesis $\left(\frac{2 \ \Delta h_{sl}}{\Delta\rho \ T_m^2}\right)$ is given the symbol **M** ($m^2 \ K^{-2} s^{-2}$) which is a material specific constant. Assuming that $T_{si} \approx T_m$, $T_{li} \approx T_m$ the thickness of a diffuse interface can now be calculated. For any given interface thickness the *driving force* *diffuseness* ($\eta_G$) may be defined as the number of pseudo atomic layers within the diffuse region of the interface which may be written as:

$$\eta_G = \frac{\zeta}{d} \tag{34}$$

where d (m) is the interplanar lattice spacing at the melting point. Dividing equation (33) on both sides by the interplanar lattice spacing and combining with equation (34) gives:

$$\eta_G = \frac{V}{G_{SLI}} \frac{1}{d\sqrt{M}} \tag{35a}$$

From equation (35a), the transition point (beyond one atomic layer thick) for the atomistically smooth to atomistically rough interface is given as:

$$\left(\frac{V}{G_{SLI}}\right) = \sqrt{M}.d \tag{35b}$$

Equation (35a) can be expressed logarithmically as:

$$\log_{10} \eta_G = \log_{10}\left(\frac{V}{G_{SLI}}\right) + \log_{10}\left(\frac{1}{d\sqrt{M}}\right) \tag{35c}$$

At the critical condition equation (35c) becomes:

$$\log_{10}(\eta_G)_C = \log_{10}\left(\frac{V}{G_{SLI}\,d}\right)_C + \log_{10}\left(\frac{1}{\sqrt{M}}\right) \tag{35d}$$

Equation (35a) can be rewritten in terms of $\dot{\varphi}_{max}$ at the critical condition as:

$$\left(\frac{V}{G_{SLI}}\right)_C = \frac{(\dot{\varphi}_{max})_C\,T_m}{G_{SLI}^2\,\Delta S_{sl}} \tag{36a}$$

Equation (35a) can further be cast in cooling rate critical dimensions (i.e. V $G_{SLI}$) as:

$$(V\,G_{SLI})_C = \frac{(\dot{\varphi}_{max})_C\,T_m}{\Delta S_{sl}} \tag{36b}$$

The *thermal diffuseness* is defined as $R_g T_m/\Delta h_{sl}$ which is the inverse of the well-known

*Jackson's criterion* [10]. The sum of the *driving force diffuseness*s and the *thermal diffuseness* ($\eta_G + \eta_\alpha$) is called the *total diffuseness* ($\eta_T$). Thus the entropy generation rate is noted to display a critical point beyond which the interface will become diffuse. The transition to an atomistically diffuse planar interface at the critical condition can be predicted from the total diffuseness as:

$$\log_{10} \eta_T = \log_{10} \left( \frac{V}{G_{SLI}\, d} \right)_C + \log_{10} \left( \frac{1}{\sqrt{M}} \right) + \log_{10} \eta_\alpha \tag{37}$$

### 2.5. Interface thickness, diffuseness and non-planar instability for binary materials

For dilute binary alloy materials the possible transitions will additionally involve diffuse interface or topographical transitions which can be topographically smooth. For a binary alloy materials, the partial derivative of the maximum entropy generation rate density with respect to the velocity while holding $\zeta$ and $C_O$ constant gives:

$$\left( \frac{\partial \dot\varphi_{max}}{\partial V} \right)_{\zeta,\, C_O} = \frac{\Delta h_{sl}\, G_{SLI}}{T_{li} \cdot T_{si}} - \frac{\Delta T_O}{D_L} \frac{V\, R_g\, ln(1/k)}{4\, m_L} \tag{38}$$

For a binary material the MEPR instability can occur when:

$$\left( \frac{\partial \dot\varphi_{max}}{\partial V} \right)_{\zeta,\, C_O} = \mathbf{0} \tag{39}$$

Equation (39) is valid at the peak of $\dot\varphi_{max}$ against velocity. Experimental comparisons show that the instability is noted at or beyond the peak. The dependence of $\dot\varphi_{max}$ on $\zeta$ and the dissipative nature of the entropy generated as a result of change of velocity are well noted in equation (31b), and is expected to oscillate the partition coefficient of the solute in the liquid. An effective partition coefficient $k_{eff}$ (dimensionless) can be inferred by comparing the peak condition to the experimental breakdown condition.

$$\left( \frac{V}{G_{SLI}} \right)_C = \frac{D_L}{\Delta T_O} \frac{2\, m_L\, \Delta h_{sl}}{T_m^2\, R_g\, ln(1/k_{eff})} \tag{40}$$

Note that, $\left( \frac{\partial^2 \dot\varphi_{max}}{\partial V^2} \right)_{\zeta,\, C_O}$, is negative for a maximization condition. Although $T_{si}$ and $T_{li}$ are unknown based on equations (32 and 38) for binary materials, the thickness of the *diffuse interface* can be approximated for dilute solutions by assuming that $T_{si} \approx T_m$ and $T_{li} \approx T_m$ to give:

$$\zeta = \frac{V}{G_{SLI}} \frac{1}{\sqrt{M-B}} \tag{41a}$$

$$\zeta = \frac{V}{G_{SLI}} \frac{1}{\sqrt{N}} \tag{41b}$$

where $N$ (m$^2$ K$^{-2}$s$^{-2}$) is defined as $\left[\left(\frac{2\ \Delta h_{sl}}{\Delta \rho_k\ T_m^2}\right) - \left(\frac{V\ \Delta T_O\ R_g \ln\left(\frac{1}{k_{eff}}\right)}{2\ G_{SLI}\ D_L\ \Delta \rho_k\ m_L}\right)\right]$, $\mathbf{M}$ (m$^2$ K$^{-2}$s$^{-2}$) is

defined as $\left(\frac{2\ \Delta h_{sl}}{\Delta \rho_k\ T_m^2}\right)$ and $\mathbf{B}$ (m$^2$ K$^{-2}$s$^{-2}$) is defined as $\left(\frac{V\ \Delta T_O\ R_g \ln\left(\frac{1}{k_{eff}}\right)}{2\ G_{SLI}\ D_L\ \Delta \rho_k\ m_L}\right)$. It is logical to assume that at least two interface layers are required to label an interface as diffuse i.e.

$$\eta_G \geq 2 \tag{42}$$

Substituting equation (34) into equation (41b) now gives the *driving force diffuseness* for a binary alloy material as:

$$\eta_G = \frac{V}{G_{SLI}} \frac{1}{d} \frac{1}{\sqrt{N}} \tag{43}$$

Taking the logarithm on both sides of equation (43) gives:

$$\log_{10} \eta_G = \log_{10}\left(\frac{V}{G_{SLI}\ d}\right) + \log_{10}\left(\frac{1}{\sqrt{N}}\right) \tag{44}$$

With a diffuse interface, the interface thickness for diffuseness instability can be obtained from equation (40) and equation (41) to give:

$$\zeta_C = \left(\frac{V}{G_{SLI}}\right)_C \frac{\sqrt{\Delta \rho_k}\ T_m}{\sqrt{\Delta h_{sl}}} \tag{45}$$

where $\zeta_C$ (m) is the critical *diffuse interface thickness* at breakdown for this possible configuration. From equation (45) the thickness is now written as:

$$(\eta_G)_C = \left(\frac{V}{G_{SLI}d}\right)_C \frac{\sqrt{\Delta\rho_k}\ T_m}{\sqrt{\Delta h_{sl}}} \tag{46a}$$

$$(\eta_G)_C = \left(\frac{V}{G_{SLI}}\right)_C \frac{1}{(\sqrt{N})_C \cdot d} \tag{46b}$$

$$\log_{10}(\eta_G)_C = \log_{10}\left(\frac{V}{G_{SLI}\cdot d}\right)_C + \log_{10}\left(\frac{\sqrt{\Delta\rho_k}\ T_m}{\sqrt{\Delta h_{sl}}}\right)_C \tag{46c}$$

The transition from an atomistically smooth to atomistically rough interface occurs when:

$$\left(\frac{V}{G_{SLI}}\right)_C = \left(\sqrt{N}\right)_C \cdot d \tag{47}$$

Equation (46) can thus also be used to infer that the *diffuse interface may persist* for a topographical instability (discussed more in section 2.6 below) and can also indicate the numerical value for the number of pseudo atomic-layers at the instability conditions. It should be remembered that a diffuse interface is associated with various fractions of solid and liquid. Note that the diffuseness at an interface is also influenced by the *thermal diffuseness* $\eta_\alpha$ (dimensionless) which may be thus connected to the formation of macroscopic smoothness and associated roughness. The equation (46) can be written for the *total diffuseness* at instability conditions as:

$$(\eta_T)_C = \left(\frac{V}{G_{SLI}\cdot d}\right)_C \frac{\sqrt{\Delta\rho}\ T_m}{\sqrt{\Delta h_{sl}}} + \eta_\alpha \tag{48a}$$

$$\log_{10}(\eta_T)_C = \log_{10}\left(\frac{V}{G_{SLI}\cdot d}\right)_C + \log_{10}\left(\frac{\sqrt{\Delta\rho}\ T_m}{\sqrt{\Delta h_{sl}}}\right)_C + \log_{10}\eta_\alpha \tag{48b}$$

From known V/G$_{SLI}$ ratios and driving force diffuseness, the instability for binary materials can be expressed in the following ways as:

$$\left(\frac{V}{G_{SLI}}\right)_C = \frac{2}{\Delta\rho_k}\left(\frac{\varphi_{max}}{N\ G_{SLI}^2}\right)_C \tag{49}$$

Equation (49) offers a sufficient condition for the onset of instability condition as described further below in the discussion section. Because this condition is based on the comparison of the entropy rate maximization it may also be recast in terms of the cooling rate (VG$_{SLI}$)$_C$:

$$(VG_{SLI})_C = \frac{2\,(\dot{\varphi}_{max})_C}{\Delta\rho_k\,N_C} \tag{50}$$

Equation (49) may also be written in terms of the number of pseudo planes:

$$(V/G_{SLI})_C = \left(\eta_G\sqrt{N}\right)_C \cdot d \tag{51a}$$

$$(VG_{SLI})_C = \sqrt{N}\,(\eta_G G_{SLI}^2)_C \cdot d \tag{51a}$$

Equations (33-51) requires that the value of $N$ to be positive so as to not violate the Second law. The implications of a negative temperature gradient are discussed below in section 3.

## 2.6 Entropy generation rate by a wave-like non-planar shape with a diffuse interface.

A non-planar topography additionally includes entropy generation terms from a configurational change when the solid and liquid fractions are rearranged [9, 23]. Additionally, a non-planar topography can exist also with a *diffuse interface*. Although rigorous details of this assessment are left for a future study, a preliminary model with *two* typical waveforms that approximate perturbations or a cellular topography are discussed in this article. For simplicity, a single harmonic is considered. The perturbation of a moving planar SLI can be described by a time independent sine wave or sine-squared expressed respectively with the diffuseness. Consider the two waveforms (shown in figure 1) described as:

$$y(x) = \varepsilon\,\sin\left(\frac{2\pi}{\lambda}x\right) \tag{52a}$$

$$y(x) = \varepsilon\,\sin^2\left(\frac{2\pi}{\lambda}x\right) \tag{52b}$$

where the $y$ direction is normal to the planar interface, the $x$ direction is along the planar interface and $\varepsilon$ is the maximum amplitude (at steady state) and $\lambda$ *is* the wavelength. It is assumed that for a fixed solidification velocity and temperature gradient, the interface thickness reaches a maximum with velocity as shown in figure 1. The amount of diffuseness at any location along the x axis is given by equations (33-35). For a perturbed interface especially with very small amplitudes, the thickness $\zeta$ is expected to achieve a minimum and a maximum at different locations on the curve. Both waveforms show an interface where the minimum value of $\zeta$ occurs at the apex of the wave growing into the liquid at a temperature $T_{li}$. The $T_{li}$ corresponds to the $\lambda/4$ position. By calculating $\zeta$ with equation 35, this yields the critical condition as:

$$\zeta_C \leq \left(\frac{V}{G_{SLI}}\right)_C M \tag{53}$$

For perturbation where the maximum amplitude occurs between $0$ and $\lambda/4$, the interface thickness is $2\zeta_C$ at $\lambda/8$, which may be expressed for the critical conditions as:

$$2\zeta_C \leq \left(\frac{V}{G_{SLI}}\right)_C M \tag{54}$$

Combining equations (53) and (54) yields the two bounds for the critical parameter for cellular shapes.

$$(\eta_G)_C \frac{d}{M} \leq \left(\frac{V}{G_{SLI}}\right)_C \leq 2(\eta_G)_C \frac{d}{M} \tag{55}$$

Alternatively, this can be written in terms of the regime for maximum entropy generation density rate in the SLI for cellular approximations.

$$\frac{\dot{\varphi}_{max} T_{si} T_{li}}{G_{SLI}^2 \Delta h_{sl}} \leq \left(\frac{V}{G_{SLI}}\right)_C \leq \frac{2 \dot{\varphi}_{max} T_{si} T_{li}}{G_{SLI}^2 \Delta h_{sl}} \tag{56}$$

A recognition of this type of bounds becomes important, as discussed below, for comparing the entropy generation rate density for atomistically planar or atomistically diffuse planar and the diffuse non-planar shape. The diffuse non-planar will additionally contain the configurational entropy terms, omitted in this article but discussed in references [9, 23].

### 3. Results and discussion

The first and second deriv*atives w.r.t to* V *at constant ζ and $G_{SLI}$* of equation (31b) indicates that the entropy generation rate will increase with velocity (equation 33) *unless* solute partitioning into the liquid is possible (equation 32). When solute partitioning is possible, the entropy rate generation term indicates a maximum, when plotted as a function of velocity (equation 38-40). As long as no other interface configuration is feasible (ones that display a higher entropy rate generation e.g. a jagged interface), the interface will remain planar during growth. Note that for $\dot{\varphi}_{max}$ it cannot be less than zero (Second Law of Thermodynamics). This implies that regardless of the sign of $G_{SLI}$ the critical $\dot{\varphi}_{max}$, can only have minimum value of zero for a planar interface. Thus a non-planar shape can always overtake a plane front morphology for a negative temperature gradient or in other words $G_{SLI}<0$ will always imply a breakdown into cells or other patterns. Additionally, because cellular shapes with a diffuse interface are seemingly restricted by the bounds of entropy from the diffuseness (equation 56), any other shape which offers an additional configurational entropy production rate increase because of complex features (e.g. dendrites) which will always emerge unless a very wide diffuse interface is possible with no partitioning.

All the interface transitions that occur, at any length scale of study, are discussed below for their dependence on $V/G_{SLI}$ (or the cooling rate V.$G_{SLI}$) and the composition, by comparing the respective entropy generation rates. The MEPR model is able to test both microscopic and topographical transitions simultaneously. For the facet to non-facet transition (*f/nf*) the change at the interface is microscopic and therefore the appropriate length for normalization is the interplanar spacing. Equation (49) is also able to predict atomistically smooth to atomistically rough interface transitions. This condition is associated with the minimum interplanar spacing in the growth plane possible i.e. when $\eta_G$ is equal to one, which becomes the transition feature from atomistically smooth to rough interface. For the instability that describes the possible onset of non-planar morphologies, the relevant length scale for normalization is offered by the diffusion length in the liquid (equation 43).

### 3.1. Pure Materials

The MEPR model is able make predictions for interface thickness and driving force diffuseness (from the imposed velocity or cooling rate) as a function of $V/G_{SLI}$ or $VG_{SLI}$, the cooling rate. The model predictions for interface thickness and diffuseness as a function of the $V/G_{SLI}$ ratio are shown in figures 2 and 3 respectively for various materials. Note that the slope is proportional to $1/\sqrt{M}$, where M is a material constant mostly determined from experiment. Figure 4 shows the plot of equation (36b) i.e. of the maximum entropy generation rate density with $VG_{SLI}$ (cooling rate). A linear relationship is seen with a slope equal to the normalized entropy of transformation (the same as the Jackson criterion). The criterion for smooth to rough interface occurs beyond a single atomic spacing which is given in equation (35b). Topographical perturbations of an interface may be of the faceted kind or smooth. The transition to a topographically jagged interface generally requires that the interface remain atomistically smooth yet become non-planar (equation 35). In conventional models this happens with anisotropy in the surface energy (that is when the second derivative of surface energy with orientation becomes significant). This is because of the fact that for any interface region, when non-planar, will provide an additional configurational entropy increase [9], which we infer that an atomistically smooth interface will always be subject to a jagged topographical instability. However, if the diffuse condition is able to provide more entropy generation than a jagged topology by additional diffuseness, then an interface can remain planar as long as diffuseness is allowed.  As diffuseness is also possible by the thermal roughening mechanism in addition to driving force induced diffuseness for non-planar, one notes that even the low melting organic materials like salol can display curved non- planar topography during growth and succinonitrile will always show a curved non- planar topography simply because of thermal roughening.

The model results for salol are shown in figure 5 which also shows the positioning of various experimentally noted microstructure patterns [40] for various growth conditions. In figure 5, experimental positions for pure salol for a facet and/or non-facet microstructural regime prediction are shown - the horizontal dotted line is a separation line which separates a facet and non-facet morphology based on the "$d$" spacing at the melting point in the <110> direction. The experimental $V/G_L$ and microstructure above the horizontal dotted line in figure 5 show a non-faceted (*nf*) wavy morphology whiles the experimental points below it show a faceted (*f*) morphology during solidification. The points formed around and close to the horizontal dotted-red-line (border line) have the potential to form facet or non-facet morphologies depending on the growth velocity, temperature gradient and crystallographic direction chosen by the interface. It is likely that the transition could initially require a short burst of extra entropy generation more than either steady state would require [9] but this is left to a further study.

According to the MEPR model, a perturbation with non-facet morphology during growth will be observed when the pseudo number of planes, $\eta_T > 1$ (or between 1 and 2). A perturbation which is related to a facet morphology is likely to be observed when $\eta_T < 1$. Figures 5 provides a visual explanation of how salol may transition from facet morphology to non-facet morphology with increasing velocity. This is an example of the effect of *driving force diffuseness* predicted theoretically by Cahn [7] and the MEPR model. Such transitions in many materials have been recorded [41-48]. The Cahn model [7] which showed for the first time that diffuseness was a function of velocity was unable to make clear quantitative predictions for the onsets of facets. The MEPR model shows how both the velocity transition predicted by Cahn [7] as well as the roughening ideas formulated by Jackson [10], may be related to the *diffuseness* and to the *topography*, thus clarifying the dependence of the *f-nf* transition on the temperature gradient. Although there are only a few experimental studies on the factors that influence *f-nf* transitions, it has been noted that both the temperature gradient and transformation velocity play a major role for such a transformation [49]. Pure bismuth, salol, germanium, benzyl, silicon, water etc., [50, 51] have the ability to exhibit both faceted and non-faceted morphologies at different crystallographic orientations and undercooling (or temperature gradients). It has experimentally been seen that at a low undercooling, hopper crystals are observed for bismuth with a faceted morphology. These experimental observations appear to be in agreement with the predictions made by equation (37). Equation (37) shows that the *f-nf* transition is dependent on the temperature gradient *and* velocity. In addition, the slope based on experimentally determined $\Delta S/T_m$ should approximately be in the order of $10^3$. The model appears to confirm for most materials.

The value of $\sqrt{M}$ is greater than one for most materials and less than one for high density materials such as osmium (0.869 *m/Ks*) and iridium (0.695 *m/Ks*). Bismuth ($\sqrt{M}$ = 0.423 *m/Ks*) and Germanium ($\sqrt{M}$ = 0.579 *m/Ks*) show the lowest values of $\sqrt{M}$ which is due to their high melting temperature, heat of fusion and the very low shrinkage noted during solidification. For polymeric materials such as succinonitrile, salol, thymol etc. the value of $\sqrt{M}$ is in the order of 10-100 *m/Ks*. In the next section the importance $\sqrt{(M-B)}$ is discussed. There is no solution possible when this number is negative. The value for M influences this aspect. Note that this number is particularly important for plastic materials like Succinonitrile and its dilute alloys for understanding the reasons for the observance of curved non-planar interface configurations when comparing equations 35 and 52-56, although as per the Jackson criterion this material could be considered as growing with facets.

### 3.2. Binary Alloys

The MEPR model shows that the *diffuse interface thickness* of a binary material may be calculated with the $V/G_{SLI}$ ratio, equation (41). It is possible as discussed further below and in the tables 1 and 2 that an effective partition coefficient may be required for accurately describing the solute gradient with a diffuse interface, one that changes with diffuseness. The *diffuse interface thickness* becomes zero when the $V/G_{SLI}$ ratio is zero. Figures 6 and 7 show the plot of thickness of the interface or number of pseudo-layers as a function of $V/G_{SLI}$ or $V.G_{SLI}$ i.e. equation (41), at a fixed solute composition and partition coefficient. Note that an exponential like behavior is observed terminating at the point where M=B i.e. when *N* approaches zero which is the limit of the *diffuse interface thickness* formulation. The growth of the interface can be steady when *N* is greater than one. However, as the *diffuse interface thickness* is subjected to high velocities the slope of the curve changes quickly when *N* becomes less than one. Note from equation (41) that the diffuse interface thickness becomes zero only at a zero velocity. When the temperature gradient is zero, the *diffuse interface thickness* becomes undefined. When **B** is equal to **M**, then **N** is zero and, $\zeta$ and $\dot{\phi}_{max}$ are both undefined. From the transition instability criterion defined by equation (39), the peak for $\dot{\phi}_{max}$ against velocity occurs when **M/B** (dimensionless) is equal to **2** i.e. **M/N$^{0.5}$** is equal to $\frac{2\sqrt{\Delta h_{sl}}}{T_m\sqrt{\Delta\rho_k}}$ (m K$^{-1}$s$^{-1}$). Further from equation (41), when **M>B** then the number of pseudo-atomic layers present within the diffuse interface region are easily related to the *driving force diffuseness* given in equation (34) in an almost linear manner. Note that the deviation from linearity sets in at a lower *V/G$_{SLI}$* as the concentration is increased.

At the condition where ***M≥N*>1**, noted from figure 6, a steady slope is observed where the *V/G$_{SLI}$* ratio shows a strong effect on the number of pseudo atomic-spacings. As the condition for ***I>N*>0** is encountered, see figure 6, only a small change in the *V/G$_{SLI}$* ratio can lead to a rapid change in the number of pseudo atomic-spacings at the interface.

The horizontal dotted-red line in figures 6 and 8 corresponds to a single atomic layer of the material formed at the interface as predicted by equation (47). The materials that solidify above the horizontal red-dotted line in figures 6 and 8 are expected to display the presence of atomistically rough interface features. Solidification below the horizontal red-dotted line indicates atomistically smooth interface. When ***B*** becomes greater than or equal to ***M***, then ***N*** is either zero or negative, and the *interface diffuseness* becomes undefined. The maximum entropy generation rate density increases with the corresponding increase in *diffuse interface thickness* and falls only when the parameter ***B*** approaches half of ***M.*** This feature of maximization indicates where instability to a non-planar topography may initiate.

Several historical experiments in gravity and microgravity conditions have shown that the critical *V/G$_L$* is a function of composition for many binary materials. Figure 8 and 9 compare the model predictions from driving force diffuseness and from total diffuseness as a function of $V/G_{SLI}$. In figure 8 is the model result for the calculated *driving force diffuseness* from experimental measurements against experimental $V/(G_{SLI}\,d)$ ratio at the critical condition based on equation (46). Tables 1 and 2 compare the experimental match with CUT, LST and MEPR models with and without the effective partition coefficient values. The predicted diffuseness is also listed. In the phase field literature the number of pseudo atomic layers in a diffuse region [52], can vary between 2 and 2750 lattice spacings which is usually an a priori assumption made of the interface thickness. From the graphs in figures 8 and 9, the diffuse interface is approximately noted to be of small to 834 lattice spacings. The calculated driving force diffuseness thickness is given in table 2 for all the alloys reported in this article. The relationship between total diffuseness and the ratio of the critical velocity $(V)_C$, to the temperature gradient $(G_{SLI})_C$ should yield a straight line as per eqn (48b) irrespective of material parameters for any growth direction (or any crystal plane spacing normal to a growth direction). The calculated total diffuseness for each binary material for this figure is given in table 2.

The model result given in figure 8 satisfies the predictions made in figures 6 and 7. For all metallic materials only one slope (equal to 0.72995 *Ks/m*) is observed. Also for plastic materials in the region below the dashed line, i.e. the atomistically smooth region, only one slope (equal to 0.07373 *Ks/m*) is observed. The implications of this are not yet fully understood in terms of diffuseness but it appears to indicate validity for the MEPR model. It is possible that this curve may indicate a basis for an effective partition ratio based on interface thickness, but this is left to future studies. For several materials like the Al alloys and Pb-Sn alloys the extent of the diffuse interface is large i.e. contains many pseudo atomic-layers. The high interface thickness calculated alloys materials are perhaps not unusual. Experimental evidence of large *interface thickness* as thick as 1 micron in size has been reported in Al-Cu alloys [53].

The influence of composition is highlighted in figure 7 where the model prediction for Al-Cu binary alloys is plotted for compositions spanning four orders of magnitude in the dilute concentration range. The model prediction shown in figure 10, shows the relationship between the calculated maximum entropy generation rate density and $V/G_{SLI}$, for different classes of binary materials. Figure 10 displays the typically noted symmetric parabolic profile of the entropy generation rate with increasing $V/G_{SLI}$. The maximum entropy generation rate density reaches a peak value and falls because the solute gradient in the liquid region begins to create new entropy compared to the amount being created in the SLI. Also note that the maximum entropy generation rate density cannot be negative and can approach a zero value only at zero $V/G_{SLI}$. Figure 11 shows the symmetric parabolic profile at low solute concentrations. Note that as expected, an indefinitely increasing entropy generation with a linear relationship to $V/G_{SLI}$ is observed at extremely dilute solute concentration (similar to figure 2). Thus at extremely low solute concentration the parameters **M** and **N** become approximately equal which reduces eqn (32) to that of a pure material when the partition ratio is one. Such a change in the partition coefficient is sometimes noted for rapid solidification conditions. The implications for the very high velocity solidification conditions were also discussed in reference 9. At very low solute concentrations, the value of **N** becomes approximately equal to **M** and the number of pseudo atomic-layers at the interface increases linearly and indefinitely as the $V/G_{SLI}$ ratio changes. Thus no other shape is able to substitute for the planar interface.

From figure 9, the calculated *total diffuseness* and the experimental measurements is plotted against the experimental $(V/G_L d)_C$ at breakdown conditions with all points labelled as either facet or non-facet as according to equation (48). The horizontal dotted-red line again serves as the transition zone between the two regimes and represents a single atomic layer for the smallest interplanar spacing growing along a selected crystallographic plane. The materials that fall above the dotted-red horizontal line are materials that show a non-facet morphology during interface breakdown. The materials that fall below the dotted-red horizontal line show a facet morphology during interface breakdown. Figure 9 also shows that one common line can be established in the non-faceted regime whereas the absence of a common line in the faceted morphological regime may be an indication of a high effect of anisotropy. It may therefore further be inferred that equation (48) holds across all velocities and gradients for any planar interface. It can be seen in figure 9 that the data points for binary materials such as SCN-Ace and SCN-Sal are below the dotted-red line which is an indication of facet morphology at interface breakdown. However, The SCN-Ace and SCN-Sal which are typically plastic crystals are not made diffuse by the driving force but only by the *thermal diffuseness*. At this condition the *thermal diffuseness* becomes the sole determinant of the interface morphology during non-planar breakdown of SCN-Ace and SCN-Sal materials. The rest of the binary materials (Al-Cr, Al-Cu, Al-Ti, Al-Zn, Pb-Ag, Pb-Bi, Pb-Sb, Pb-Sn, and Sn-Pb) show non-facet morphology while Bi-Sn will display facet morphology at breakdown which is in agreement with all experimental observations. Figure 9 further shows that a transition from facet to non-facet transition is highly probable for certain alloys predicted by the MEPR model depending of the solidification conditions. There is currently a paucity of experimental data regarding facet transformations for binary materials with the exception of $Al_2O_3$-MgO [49]. For the $Al_2O_3$-MgO study [49], a laser surface scanning technique [54] was employed for independent control of the velocity and gradient. In this experiment (i.e., for $Al_2O_3$-MgO) [49], the transition from facet to non-facet and again to a facet state was reported. Similar results have also been noted earlier by Jackson and Miller [46] in undercooled alloys for hexachloroethane and ammonium chloride; by Glicksman and Schaeffer [43] for white phosphorus; and for aperiodic (quasicrystalline) phases in the Pb-Bi and Cu-Sn systems [55, 56]. Similarly the observation of a facet-to-non-facet (*f-nf*) transition for Al-Ti, SCN-Sal and SCN-Ace materials at an increased velocity can be explained again in accordance with the experimental observation [7, 40]. Note that the facet (jagged topography) is seen sometimes in preference to a diffuse interface condition and multiple transitions are possible.

It is noted that when the maximum entropy generation rate density is plotted against the interface thickness, equation (31) (figure 12) an asymmetric bell shaped curve is seen for binary material. Without further comment, we note that the shape of this curve is similar to the LST predictions for plot of perturbation wavelength and imposed solidification conditions.

The maximum entropy generation rate density displays a diminishing peak height and size with an increase in the solute concentration as shown for Al-Cu in figure 13. Further in figure 13, it is noted that the entropy vs. the interface thickness curve flattens for very dilute solute concentrations. This happens at $M >> B$, where the effect of solute diffusivity and partition coefficient in the liquid become of low significance. It may be inferred that at a high maximum entropy generation rate density, the partition coefficient could increase to accommodate the increase in velocity and/or number of pseudo atomic layers. An effective partition coefficient for a number of binary materials is calculated using the peak with the experimental reported measurements. Table 2 lists the equilibrium partition numbers and the effective numbers based on the comparison.

## 4. Summary and conclusion

The key MEPR condition for interface diffuseness or topographical change is primarily related to the maximum entropy rate and thus related to the composition, velocity of solidification, the temperature gradient encountered in the solid-liquid zone, and the *effective* partition coefficient when a solute gradient in the liquid is established. The MEPR model postulates that that entropy generation is maximized when an interface transition occurs to a different configuration whether an atomistic or a topographical variant. The model for pure and binary materials is able to quantitatively predict the size of a diffuse interface and the number of pseudo-atomic layers present. A comparison with historically experimentally measured breakdown shows that the model is also able to account for the interface topography as being either facet or non-facet kind. The model also appears to correctly predict an explanation for the transition from facet to non-facet (*f/nf*) planar or non-planar topography as dependent on velocity and the temperature gradient. The MEPR predictions compare reasonable with the reported experimental measurements for over ninety binary material compositions. The new criterion may allow for a better estimate of the *solute diffusion constant* in binary alloys than that available previously from solidification measurements [23] and relating to the CUT or LST models. It is possible that the CUT and LST criteria for interface instability may only be necessary conditions, but not sufficient enough to describe comprehensive interface instability criterion applicable to all material types and across all possible interface configurations that arise from atomistic or configurational variants.

**Acknowledgements and Funding Sources**

This article summarizes research performed for the Ph.D. award for Yaw Delali Bensah. Funding for the research came from his personal savings and partial support from MHI Inc. Cincinnati Ohio, USA. A partial tuition scholarship support received from University of Cincinnati, Cincinnati, Ohio, for Yaw Delali Bensah is also acknowledged. Professor J. A. Sekhar acknowledges funding from MHI Inc. and the Institute of Design and Thermodynamics, Cincinnati, Ohio, USA.

Table 1. A summary of $V/G_L$ at instability conditions for experimental breakdown compared with, CUT and LST. The coefficients of diffusion given are for independent experimentally measured from different authors and are corrected to their solute concentrations at the solidus temperatures. The constants $Z_{CUT}$ and $Z_{LST}$ are deviations from $(V/G_L)_{exp}$ for CUT and LST criterion respectively. Experimental data is individually referenced in [23].

| Binary material | $D_L \times 10^{-9}$ ($m^2s^{-1}$) at $T_S$ | $(V/G_L)_C$ ratios at breakdown ($\times 10^{-9}$) ($m^2K^{-1}s^{-1}$) | | | $Z_{CUT}$ (*dimensionless*) | $Z_{LST}$ (*dimensionless*) |
| | | Expt | CUT | LST | | |
|---|---|---|---|---|---|---|
| Al-0.102 *wt%*Cr | 0.26051 | 29.760 | 3.7912 | 2.0909 | 7.8496 | 14.232 |
| Al-0.102 *wt%*Cr | 0.26051 | 24.596 | 3.7912 | 2.0909 | 6.4877 | 11.763 |
| Al-0.201 *wt%*Cr | 0.26115 | 11.962 | 1.9277 | 1.0632 | 6.2051 | 11.251 |
| Al-0.201 *wt%*Cr | 0.26115 | 11.565 | 1.9277 | 1.0632 | 5.9994 | 10.878 |
| Al-0.328 *wt%*Cr | 0.26198 | 9.766 | 1.1844 | 0.6532 | 8.2458 | 14.951 |
| Al-0.328 *wt%*Cr | 0.26198 | 9.276 | 1.1844 | 0.6532 | 7.8324 | 14.201 |
| Al-0.328 *wt%*Cr | 0.26198 | 8.301 | 1.1844 | 0.6532 | 7.0086 | 12.708 |
| Al-0.328 *wt%*Cr | 0.26198 | 9.359 | 1.1844 | 0.6532 | 7.9026 | 14.328 |
| Al-0.328 *wt%*Cr | 0.26198 | 8.912 | 1.1844 | 0.6532 | 7.5245 | 13.643 |
| Al-0.328 *wt%*Cr | 0.26198 | 7.541 | 1.1844 | 0.6532 | 6.3672 | 11.545 |
| Al-0.025 *wt%*Cu | 7.4519 | 23.913 | 12.166 | 6.7097 | 1.9656 | 3.5639 |
| Al-0.025 *wt%*Cu | 7.4519 | 41.026 | 12.166 | 6.7097 | 3.3723 | 6.1144 |
| Al-0.47 *wt%*Cu | 7.1474 | 12.069 | 0.6191 | 0.3414 | 19.4952 | 35.3474 |
| Al-0.2 *wt%*Cu | 7.3318 | 4.8 | 1.4947 | 0.8244 | 3.2114 | 5.8227 |
| Al-0.73 *wt%*Cu | 6.9709 | 1.1 | 0.3882 | 0.2141 | 2.8339 | 5.1382 |
| Al-0.024 *wt%*Ti | 2.0392 | 4.393 | 3.0123 | 1.6614 | 1.4772 | 2.6439 |
| Al-0.054 *wt%*Ti | 2.0402 | 1.382 | 1.3393 | 0.7387 | 0.6565 | 1.8715 |
| Al-0.083 *wt%*Zn | 4.4419 | 42.3 | 25.444 | 14.033 | 1.6633 | 3.0158 |
| Al-0.083 *wt%*Zn | 4.4419 | 40.0 | 25.444 | 14.033 | 1.5721 | 2.8504 |
| Al-0.083 *wt%*Zn | 4.4419 | 37.8 | 25.444 | 14.033 | 1.4869 | 2.6961 |
| Al-0.096 *wt%*Zn | 4.4399 | 24.9 | 21.987 | 12.127 | 1.1310 | 2.0507 |
| Al-0.096 *wt%*Zn | 4.4399 | 27.9 | 21.987 | 12.127 | 1.2699 | 2.3026 |
| Al-0.096 *wt%*Zn | 4.4399 | 26.7 | 21.987 | 12.127 | 1.2151 | 2.2031 |
| Al-0.375 *wt%*Zn | 4.3983 | 6.53 | 5.5667 | 3.0702 | 1.1725 | 2.1258 |
| Al-0.375 *wt%*Zn | 4.3983 | 7.20 | 5.5667 | 3.0702 | 1.2939 | 2.3460 |
| Al-0.375 *wt%*Zn | 4.3983 | 7.71 | 5.5667 | 3.0702 | 1.3845 | 2.5103 |

Table 1(continued). A summary of $V/G_L$ at instability conditions for experimental breakdown compared with, CUT and LST. The coefficients of diffusion given are for independent experimentally measured from different authors and are corrected to their solute concentrations at the solidus temperatures. The constants $Z_{CUT}$ and $Z_{LST}$ are deviations from $(V/G_L)_{exp}$ for CUT and LST criterion respectively. Experimental data is individually referenced in [23].

| Binary material | $D_L \times 10^{-9}$ (m²s⁻¹) at $T_S$ | $(V/G_L)_C$ ratios at breakdown ($\times 10^{-9}$) ($m^2 K^{-1} s^{-1}$) | | | $Z_{CUT}$ (dimensionless) | $Z_{LST}$ (dimensionless) |
|---|---|---|---|---|---|---|
| | | Expt | CUT | LST | | |
| Bi-0.057 wt% Sn | 2.4954 | 0.615 | 0.3396 | 0.4057 | 1.8107 | 1.5156 |
| Bi-0.571 wt% Sn | 1.5899 | 1.176 | 0.0216 | 0.0258 | 54.3733 | 45.5143 |
| Pb-0.0001 wt% Ag | 6.3919 | 355.319 | 281.86 | 177.17 | 1.2614 | 2.0068 |
| Pb-0.00025 wt% Ag | 6.3915 | 162.338 | 112.74 | 70.862 | 1.4409 | 2.2924 |
| Pb-0.0005 wt% Ag | 6.3908 | 62.037 | 56.361 | 35.427 | 1.1014 | 1.7523 |
| Pb-0.00075 wt% Ag | 6.39 | 36.8 | 37.569 | 23.615 | 0.9801 | 1.5593 |
| Pb-0.0001 wt% Ag | 5.8678 | 355.319 | 258.578 | 162.535 | 1.3741 | 2.1861 |
| Pb-0.00025 wt% Ag | 5.8677 | 162.338 | 103.429 | 65.0125 | 1.5685 | 49.5875 |
| Pb-0.0005 wt% Ag | 5.8675 | 62.037 | 51.7126 | 32.5051 | 1.1989 | 49.9521 |
| Pb-0.00075 wt% Ag | 5.8672 | 36.8 | 34.4738 | 21.6693 | 1.0668 | 49.9653 |
| Pb-0.0089 wt% Sb | 2.9472 | 27.0 | 58.751 | 36.929 | 0.4593 | 0.7307 |
| Pb-0.0179 wt% Sb | 2.9460 | 13.369 | 29.255 | 18.389 | 0.4567 | 0.7266 |
| Pb-0.0179 wt% Sb | 2.9460 | 11.546 | 29.255 | 18.389 | 0.3944 | 0.6275 |
| Pb-0.0179 wt% Sb | 2.9460 | 10.823 | 29.255 | 18.389 | 0.3697 | 0.5882 |
| Pb-0.0265 wt% Sb | 2.9449 | 7.801 | 19.704 | 12.385 | 0.3956 | 0.6294 |
| Pb-0.0354 wt% Sb | 2.9439 | 6.943 | 14.772 | 9.2854 | 0.4697 | 0.7472 |
| Pb-0.01 wt% Sn | 1.6556 | 309.259 | 76.080 | 47.822 | 3.2601 | 5.1865 |
| Pb-0.03 wt% Sn | 1.6547 | 89.634 | 25.345 | 15.931 | 2.8358 | 4.5115 |
| Pb-0.05 wt% Sn | 1.6538 | 53.261 | 15.198 | 9.5533 | 2.8095 | 4.4696 |
| Pb-0.06 wt% Sn | 1.6534 | 61.475 | 12.662 | 7.9588 | 3.8921 | 6.1920 |
| Pb-0.1 wt% Sn | 1.6516 | 47.25 | 7.5882 | 4.769 | 4.9897 | 7.9382 |
| Pb-0.15 wt% Sn | 1.6494 | 25.615 | 5.0514 | 3.1752 | 4.0616 | 6.4616 |
| Pb-0.15 wt% Sn | 1.6494 | 260.241 | 5.0514 | 3.1752 | 41.2636 | 65.6466 |
| Pb-0.15 wt% Sn | 1.6494 | 305.376 | 5.0514 | 3.1752 | 48.4202 | 77.0322 |
| Pb-0.15 wt% Sn | 1.6494 | 344.33 | 5.0514 | 3.1752 | 54.5967 | 86.8584 |
| Pb-0.15 wt% Sn | 1.6494 | 328.571 | 5.0514 | 3.1752 | 52.0980 | 82.8832 |

Table 1(continued). A summary of $V/G_L$ at instability conditions for experimental breakdown compared with, CUT and LST. The coefficients of diffusion given are for independent experimentally measured from different authors and are corrected to their solute concentrations at the solidus temperatures. The constants $Z_{CUT}$ and $Z_{LST}$ are deviations from $(V/G_L)_{exp}$ for CUT and LST criterion respectively. Experimental data is individually referenced in [23].

| Binary material | $D_L \times 10^{-9}$ ($m^2s^{-1}$) at $T_S$ | $(V/G_L)_C$ ratios at breakdown ($\times 10^{-9}$) ($m^2K^{-1}s^{-1}$) | | | $Z_{CUT}$ (*dimensionless*) | $Z_{LST}$ (*dimensionless*) |
|---|---|---|---|---|---|---|
| | | Expt | CUT | LST | | |
| SCN-0.5 *wt%* Sal | 0.395 | 0.589 | 0.1797 | 0.1789 | 3.2742 | 3.2889 |
| SCN-0.7 *wt%* Sal | 0.395 | 1.086 | 0.2516 | 0.2505 | 4.3071 | 4.3264 |
| SCN-0.7 *wt%* Sal | 0.395 | 0.589 | 0.2516 | 0.2505 | 2.3368 | 2.3473 |
| SCN-0.7 *wt%* Sal | 0.395 | 1.231 | 0.2516 | 0.2505 | 4.8769 | 4.8988 |
| SCN-0.5 *wt%* Sal | 0.690 | 0.5895 | 0.3139 | 0.3126 | 1.8744 | 1.8828 |
| SCN-0.7 *wt%* Sal | 0.690 | 1.0869 | 0.4395 | 0.4376 | 2.4656 | 2.4767 |
| SCN-0.7 *wt%* Sal | 0.690 | 0.5897 | 0.4395 | 0.4376 | 1.3378 | 1.3438 |
| SCN-0.7 *wt%* Sal | 0.690 | 1.2308 | 0.4395 | 0.4376 | 2.7918 | 2.8044 |
| SCN-0.5*wt%* Ace | 0.9552 | 0.8333 | 0.0723 | 0.0719 | 13.337 | 13.397 |
| SCN-0.1*wt%* Ace | 0.9552 | 0.6000 | 0.3615 | 0.3599 | 1.9235 | 1.9321 |
| SCN-0.1 *wt%* Ace | 0.9552 | 0.4188 | 0.3615 | 0.3599 | 1.3428 | 1.3488 |
| SCN-0.165 *wt%* Ace | 0.9552 | 0.7647 | 0.2191 | 0.2181 | 4.0439 | 4.0621 |
| SCN-0.056 *wt%* Ace[MG] | 0.9552 | 4.4400 | 0.6455 | 0.6426 | 7.9723 | 8.0080 |
| SCN-0.12 *wt%* Ace[MG] | 0.9552 | 1.2833 | 0.3012 | 0.2999 | 4.9366 | 4.9587 |
| SCN-0.106 *wt%* Ace | 0.9552 | 0.4289 | 0.3410 | 0.3395 | 1.4576 | 1.4641 |
| SCN-0.5*wt%* Ace | 1.270 | 0.8333 | 0.0961 | 0.0956 | 10.031 | 10.076 |
| SCN-0.1*wt%* Ace | 1.270 | 0.6000 | 0.4806 | 0.4785 | 1.4467 | 1.4532 |
| SCN-0.1 *wt%* Ace | 1.270 | 0.4188 | 0.4806 | 0.4785 | 1.0099 | 1.0144 |
| SCN-0.165 *wt%* Ace | 1.270 | 0.7647 | 0.2913 | 0.2899 | 3.0416 | 3.0552 |
| SCN-0.056 *wt%* Ace[MG] | 1.270 | 4.4400 | 0.8583 | 0.8544 | 5.9962 | 6.0231 |
| SCN-0.12 *wt%* Ace[MG] | 1.270 | 1.2833 | 0.4005 | 0.3987 | 3.7129 | 3.7296 |
| SCN-0.106 *wt%* Ace | 1.270 | 0.4289 | 0.4534 | 0.4514 | 1.0963 | 1.1012 |

Table 1(continued). A summary of $V/G_L$ at instability conditions for experimental breakdown compared with, CUT and LST. The coefficients of diffusion given are for independent experimentally measured from different authors and are corrected to their solute concentrations at the solidus temperatures. The constants $Z_{CUT}$ and $Z_{LST}$ are deviations from $(V/G_L)_{exp}$ for CUT and LST criterion respectively. Experimental data is individually referenced in [23].

| Binary material | $D_L \times 10^{-9}$ (m²s⁻¹) at $T_S$ | $(V/G_L)_C$ ratios at breakdown (×10⁻⁹) (m²K⁻¹s⁻¹) | | | $Z_{CUT}$ (dimensionless) | $Z_{LST}$ (dimensionless) |
|---|---|---|---|---|---|---|
| | | Expt | CUT | LST | | |
| Pb-0.1 *wt%* Bi | 1.7719 | 5.72 | 8.0419 | 2.6161 | 0.7113 | 2.1865 |
| Pb-0.2 *wt%* Bi | 1.7676 | 3.144 | 4.0113 | 1.3049 | 0.7838 | 2.4094 |
| Pb-0.3 *wt%* Bi | 1.7634 | 2.00 | 2.6677 | 0.8678 | 0.7497 | 2.3046 |
| Pb-0.1 *wt%* Bi | 2.7619 | 5.72 | 12.535 | 4.0779 | 0.4563 | 1.4027 |
| Pb-0.2 *wt%* Bi | 2.7534 | 3.144 | 6.2482 | 2.0326 | 0.5032 | 1.5468 |
| Pb-0.3 *wt%* Bi | 2.7448 | 2.00 | 4.1525 | 1.3508 | 0.4816 | 1.4806 |
| Sn-0.0024 *wt%* Pb | 1.6556 | 52.381 | 129.76 | 83.896 | 0.4037 | 0.6244 |
| Sn-0.006 *wt%* Pb | 1.6547 | 59.091 | 51.894 | 33.552 | 1.1387 | 1.7612 |
| Sn-0.015 *wt%* Pb | 1.6538 | 10.0 | 20.748 | 13.414 | 0.4819 | 0.7455 |
| Sn-0.02 *wt%* Pb | 1.6534 | 11.429 | 15.557 | 10.058 | 0.7346 | 1.1363 |
| Sn-0.02 *wt%* Pb | 1.6516 | 9.412 | 15.557 | 10.058 | 0.6050 | 0.9357 |
| Sn-0.02 *wt%* Pb | 1.6494 | 8.00 | 15.557 | 10.058 | 0.5143 | 0.7954 |
| Sn-0.0015 *wt%* Pb | 1.6494 | 152.941 | 207.62 | 134.24 | 0.7366 | 1.1393 |
| Sn-0.012 *wt%* Pb | 1.6494 | 12.6 | 25.939 | 16.771 | 0.4858 | 0.7513 |
| Sn-0.0046 *wt%* Pb | 1.6494 | 73.913 | 67.693 | 43.767 | 1.0918 | 1.6888 |
| Sn-0.012 *wt%* Pb | 1.6494 | 20.323 | 25.939 | 16.771 | 0.7835 | 1.2118 |
| Sn-0.012 *wt%* Pb | 1.6556 | 15.0 | 25.939 | 16.771 | 0.5783 | 0.8944 |
| Sn-0.012 *wt%* Pb | 1.6547 | 14.318 | 25.939 | 16.771 | 0.5520 | 0.8538 |

Table 2. A summary of the *interface thickness*, *driving force diffuseness* and *total diffuseness* obtained from the model results at instability for different materials. Also shown are the effective partition coefficient $(k_{eff})_C$ and equilibrium partition coefficient ($k$). Although $k$ is non-dimensional the numerical value depends on the concentration units chosen. The $(k_{eff})_C$ is the value of k where the peak is noted in the entropy generation vs. velocity/$G_{SLI}$ plot in figure 10. Experimental data is individually referenced in [23].

| Binary material | $\zeta_C$ (nm) | $(\eta_G)_C$ (dimensionless) | $\eta_\alpha$ (dimensionless) | $(\eta_T)_C$ (dimensionless) | $(keff)_C$ (dimensionless) | $k$ (dimensionless) |
|---|---|---|---|---|---|---|
| Al-0.102 *wt%*Cr | 12.320 | 51.9147 | 0.7414 | 52.6561 | 1.1858 | 1.3288 |
| Al-0.102 *wt%*Cr | 10.183 | 42.9074 | 0.7414 | 43.6488 | 1.2289 | 1.3288 |
| Al-0.201 *wt%*Cr | 4.9521 | 20.8679 | 0.7414 | 21.6093 | 1.2406 | 1.3288 |
| Al-0.201 *wt%*Cr | 4.7879 | 20.1762 | 0.7414 | 20.9175 | 1.2498 | 1.3288 |
| Al-0.328 *wt%*Cr | 4.0430 | 17.0383 | 0.7414 | 17.7796 | 1.1762 | 1.3288 |
| Al-0.328 *wt%*Cr | 3.8404 | 16.1841 | 0.7414 | 16.9255 | 1.1863 | 1.3288 |
| Al-0.328 *wt%*Cr | 3.4364 | 14.4819 | 0.7414 | 15.2233 | 1.2104 | 1.3288 |
| Al-0.328 *wt%*Cr | 3.8748 | 16.3291 | 0.7414 | 17.0705 | 1.1845 | 1.3288 |
| Al-0.328 *wt%*Cr | 3.6894 | 15.5478 | 0.7414 | 16.2891 | 1.1946 | 1.3288 |
| Al-0.328 *wt%*Cr | 3.1219 | 13.1565 | 0.7414 | 13.8979 | 1.2339 | 1.3288 |
| Al-0.025 *wt%*Cu | 9.8999 | 41.7138 | 0.7414 | 42.4552 | 0.4604 | 0.0939 |
| Al-0.025 *wt%*Cu | 16.9844 | 71.5649 | 0.7414 | 72.3063 | 0.6363 | 0.0939 |
| Al-0.47 *wt%*Cu | 4.9965 | 21.0750 | 0.7414 | 21.8164 | 0.9247 | 0.0939 |
| Al-0.2 *wt%*Cu | 1.9872 | 8.3765 | 0.7414 | 9.1179 | 0.6219 | 0.0939 |
| Al-0.73 *wt%*Cu | 0.4554 | 1.9220 | 0.7414 | 2.6634 | 0.5832 | 0.0939 |
| Al-0.024 *wt%*Ti | 1.8185 | 7.7379 | 0.7414 | 8.4793 | 20983.463 | 8.2993 |
| Al-0.054 *wt%*Ti | 0.5723 | 2.4352 | 0.7414 | 3.1765 | 1276033.55 | 8.2993 |
| Al-0.083 *wt%*Zn | 17.521 | 73.825 | 0.7414 | 74.567 | 0.5803 | 0.4105 |
| Al-0.083 *wt%*Zn | 16.559 | 69.776 | 0.7414 | 70.517 | 0.5623 | 0.4105 |
| Al-0.083 *wt%*Zn | 15.664 | 65.999 | 0.7414 | 66.741 | 0.5440 | 0.4105 |
| Al-0.096 *wt%*Zn | 10.295 | 43.379 | 0.7414 | 44.121 | 0.4492 | 0.4105 |
| Al-0.096 *wt%*Zn | 11.559 | 48.707 | 0.7414 | 49.449 | 0.4903 | 0.4105 |
| Al-0.096 *wt%*Zn | 11.060 | 46.603 | 0.7414 | 47.344 | 0.4747 | 0.4105 |
| Al-0.375 *wt%*Zn | 2.7020 | 11.386 | 0.7414 | 12.127 | 0.4617 | 0.4105 |
| Al-0.375 *wt%*Zn | 2.9819 | 12.565 | 0.7414 | 13.306 | 0.4965 | 0.4105 |
| Al-0.375 *wt%*Zn | 3.1907 | 13.445 | 0.7414 | 14.186 | 0.5197 | 0.4105 |

Table 2(continued). A summary of the *interface thickness*, *driving force diffuseness* and *total diffuseness* obtained from the model results at instability for different materials. Also shown are the effective partition coefficient $(k_{eff})_C$ and equilibrium partition coefficient ($k$). Although $k$ is non-dimensional the numerical value depends on the concentration units chosen. The $(k_{eff})_C$ is the value of k where the peak is noted in the entropy generation vs. velocity/$G_{SLI}$ plot in figure 10. Experimental data is individually referenced in [23].

| Binary material | $\zeta_C$ (nm) | $(\eta_G)_C$ (dimensionless) | $\eta_\alpha$ (dimensionless) | $(\eta_T)_C$ (dimensionless) | $(k_{eff})_C$ (dimensionless) | $k$ (dimensionless) |
|---|---|---|---|---|---|---|
| Bi-0.057 wt% Sn | 0.26737 | 0.6932 | 0.4007 | 1.0939 | 0.2969 | 0.0306 |
| Bi-0.571 wt% Sn | 0.45212 | 1.1729 | 0.4007 | 1.5736 | 0.9551 | 0.0306 |
| Pb-0.0001 wt%Ag | 259.982 | 901.929 | 1.0382 | 902.967 | 0.9748 | 0.0449 |
| Pb-0.00025 wt%Ag | 118.780 | 412.072 | 1.0382 | 413.110 | 0.9779 | 0.0449 |
| Pb-0.0005 wt%Ag | 45.3916 | 157.473 | 1.0382 | 158.511 | 0.9712 | 0.0449 |
| Pb-0.00075 wt%Ag | 26.9260 | 93.413 | 1.0382 | 94.450 | 0.9677 | 0.0449 |
| Pb-0.0001 wt%Ag | 259.982 | 901.929 | 1.0382 | 902.967 | 0.9768 | 0.0449 |
| Pb-0.00025 wt%Ag | 118.780 | 412.072 | 1.0382 | 413.110 | 0.9797 | 0.0449 |
| Pb-0.0005 wt%Ag | 45.3916 | 157.473 | 1.0382 | 158.511 | 0.9735 | 0.0449 |
| Pb-0.00075 wt%Ag | 26.9260 | 93.412 | 1.0382 | 94.450 | 0.9703 | 0.0449 |
| Pb-0.0089 wt% Sb | 19.7555 | 55.9591 | 1.0382 | 56.9973 | 0.0542 | 0.5727 |
| Pb-0.0179 wt% Sb | 9.7819 | 27.7078 | 1.0382 | 28.7460 | 0.0533 | 0.5727 |
| Pb-0.0179 wt% Sb | 8.4479 | 23.9295 | 1.0382 | 24.9677 | 0.0336 | 0.5727 |
| Pb-0.0179 wt% Sb | 7.9190 | 22.4312 | 1.0382 | 23.4694 | 0.0268 | 0.5727 |
| Pb-0.0265 wt% Sb | 5.7078 | 16.1678 | 1.0382 | 17.2061 | 0.0339 | 0.5727 |
| Pb-0.0354 wt% Sb | 5.0804 | 14.3905 | 1.0382 | 15.4287 | 0.05781 | 0.5727 |
| Pb-0.01 wt%Sn | 226.281 | 785.017 | 1.0382 | 786.055 | 0.8505 | 0.6364 |
| Pb-0.03 wt%Sn | 65.5840 | 227.5287 | 1.0382 | 228.566 | 0.8301 | 0.6364 |
| Pb-0.05 wt%Sn | 38.9702 | 135.1997 | 1.0382 | 136.238 | 0.8287 | 0.6364 |
| Pb-0.06 wt%Sn | 44.9807 | 156.053 | 1.0382 | 157.091 | 0.8731 | 0.6364 |
| Pb-0.1 wt%Sn | 34.5721 | 119.945 | 1.0382 | 120.983 | 0.8996 | 0.6364 |
| Pb-0.15 wt%Sn | 18.7424 | 65.0273 | 1.0382 | 66.0655 | 0.8781 | 0.6364 |
| Pb-0.15 wt%Sn | 190.415 | 660.648 | 1.0382 | 661.6865 | 0.9873 | 0.6364 |
| Pb-0.15 wt%Sn | 223.439 | 775.229 | 1.0382 | 776.267 | 0.9892 | 0.6364 |
| Pb-0.15 wt%Sn | 251.941 | 874.117 | 1.0382 | 875.155 | 0.9904 | 0.6364 |
| Pb-0.15 wt%Sn | 240.411 | 834.112 | 1.0382 | 835.150 | 0.9899 | 0.6364 |

Table 2(continued). A summary of the *interface thickness*, *driving force diffuseness* and *total diffuseness* obtained from the model results at instability for different materials. Also shown are the effective partition coefficient $(k_{eff})_C$ and equilibrium partition coefficient ($k$). Although $k$ is non-dimensional the numerical value depends on the concentration units chosen. The $(k_{eff})_C$ is the value of k where the peak is noted in the entropy generation vs. velocity/$G_{SLI}$ plot in figure 10. Experimental data is individually referenced in [23].

| Binary material | $\zeta_C$ (*nm*) | $(\eta_G)_C$ (*dimensionless*) | $\eta_a$ (*dimensionless*) | $(\eta_T)_C$ (*dimensionless*) | $(keff)_C$ (*dimensionless*) | $k$ (*dimensionless*) |
|---|---|---|---|---|---|---|
| SCN-0.5 *wt%* Sal | 0.04346 | 0.0952 | 0.7436 | 0.8388 | 0.6389 | 0.1814 |
| SCN-0.7 *wt%* Sal | 0.08014 | 0.1755 | 0.7436 | 0.9192 | 0.7117 | 0.1814 |
| SCN-0.7 *wt%* Sal | 0.04348 | 0.0952 | 0.7436 | 0.8389 | 0.5342 | 0.1814 |
| SCN-0.7 *wt%* Sal | 0.09075 | 0.1987 | 0.7436 | 0.9424 | 0.7405 | 0.1814 |
| SCN-0.5 *wt%* Sal | 0.04346 | 0.0952 | 0.7436 | 0.8388 | 0.4573 | 0.1814 |
| SCN-0.7 *wt%* Sal | 0.08014 | 0.1755 | 0.7436 | 0.9192 | 0.5520 | 0.1814 |
| SCN-0.7 *wt%* Sal | 0.04348 | 0.0952 | 0.7436 | 0.8389 | 0.3345 | 0.1814 |
| SCN-0.7 *wt%* Sal | 0.09075 | 0.1987 | 0.7436 | 0.9424 | 0.5917 | 0.1814 |
| SCN-0.5*wt%* Ace | 0.06144 | 0.1347 | 0.7436 | 0.8783 | 0.8981 | 0.1012 |
| SCN-0.1*wt%* Ace | 0.04424 | 0.0969 | 0.7436 | 0.8406 | 0.4749 | 0.1012 |
| SCN-0.1 *wt%* Ace | 0.03088 | 0.0677 | 0.7436 | 0.8113 | 0.3442 | 0.1012 |
| SCN-0.165 *wt%* Ace | 0.05638 | 0.1236 | 0.7436 | 0.8672 | 0.7018 | 0.1012 |
| SCN-0.056 *wt%* Ace[MG] | 0.32737 | 0.7175 | 0.7436 | 1.4612 | 0.8356 | 0.1012 |
| SCN-0.12 *wt%* Ace[MG] | 0.09462 | 0.2074 | 0.7436 | 0.9510 | 0.7482 | 0.1012 |
| SCN-0.106 *wt%* Ace | 0.03163 | 0.0693 | 0.7436 | 0.8129 | 0.3744 | 0.1012 |
| SCN-0.5*wt%* Ace | 0.06144 | 0.1347 | 0.7436 | 0.8783 | 0.8668 | 0.1012 |
| SCN-0.1*wt%* Ace | 0.04424 | 0.0969 | 0.7436 | 0.8406 | 0.3716 | 0.1012 |
| SCN-0.1 *wt%* Ace | 0.03088 | 0.0677 | 0.7436 | 0.8113 | 0.2422 | 0.1012 |
| SCN-0.165 *wt%* Ace | 0.05638 | 0.1236 | 0.7436 | 0.8672 | 0.6244 | 0.1012 |
| SCN-0.056 *wt%* Ace[MG] | 0.32737 | 0.7175 | 0.7436 | 1.4612 | 0.7876 | 0.1012 |
| SCN-0.12 *wt%* Ace[MG] | 0.09462 | 0.2074 | 0.7436 | 0.9510 | 0.6799 | 0.1012 |
| SCN-0.106 *wt%* Ace | 0.03163 | 0.0693 | 0.7436 | 0.8129 | 0.2708 | 0.1012 |

Table 2(continued). A summary of the *interface thickness*, *driving force diffuseness* and *total diffuseness* obtained from the model results at instability for different materials. Also shown are the effective partition coefficient $(k_{eff})_C$ and equilibrium partition coefficient $(k)$. Although $k$ is non-dimensional the numerical value depends on the concentration units chosen. The $(k_{eff})_C$ is the value of $k$ where the peak is noted in the entropy generation vs. velocity/$G_{SLI}$ plot in figure 10. Experimental data is individually referenced in [23].

| Binary material | $\zeta_C$ (nm) | $(\eta_G)_C$ (dimensionless) | $\eta_\alpha$ (dimensionless) | $(\eta_T)_C$ (dimensionless) | $(keff)_C$ (dimensionless) | $k$ (dimensionless) |
|---|---|---|---|---|---|---|
| Pb-0.1 wt% Bi | 4.1852 | 14.519 | 1.0382 | 15.557 | 0.2735 | 0.5789 |
| Pb-0.2 wt% Bi | 2.3004 | 7.9798 | 1.0382 | 9.0180 | 0.3083 | 0.5789 |
| Pb-0.3 wt% Bi | 1.4634 | 5.0759 | 1.0382 | 6.1142 | 0.2922 | 0.5789 |
| Pb-0.1 wt% Bi | 4.1852 | 14.519 | 1.0382 | 15.557 | 0.1325 | 0.5789 |
| Pb-0.2 wt% Bi | 2.3004 | 7.9798 | 1.0382 | 9.0180 | 0.1599 | 0.5789 |
| Pb-0.3 wt% Bi | 1.4634 | 5.0759 | 1.0382 | 6.1142 | 0.1473 | 0.5789 |
| Sn-0.0024 wt% Pb | 16.972 | 63.075 | 0.5932 | 63.668 | 0.0304 | 0.1547 |
| Sn-0.006 wt% Pb | 19.146 | 71.155 | 0.5932 | 71.748 | 0.2899 | 0.1547 |
| Sn-0.015 wt% Pb | 3.2400 | 12.042 | 0.5932 | 12.635 | 0.0537 | 0.1547 |
| Sn-0.02 wt% Pb | 3.7029 | 13.762 | 0.5932 | 14.355 | 0.1467 | 0.1547 |
| Sn-0.02 wt% Pb | 3.0494 | 11.333 | 0.5932 | 11.926 | 0.0973 | 0.1547 |
| Sn-0.02 wt% Pb | 2.5920 | 9.6333 | 0.5932 | 10.226 | 0.0645 | 0.1547 |
| Sn-0.0015 wt% Pb | 49.553 | 184.17 | 0.5932 | 184.76 | 0.1475 | 0.1547 |
| Sn-0.012 wt% Pb | 4.0824 | 15.172 | 0.5932 | 15.766 | 0.0549 | 0.1547 |
| Sn-0.0046 wt% Pb | 23.948 | 89.003 | 0.5932 | 89.597 | 0.2749 | 0.1547 |
| Sn-0.012 wt% Pb | 6.5846 | 24.472 | 0.5932 | 25.065 | 0.1654 | 0.1547 |
| Sn-0.012 wt% Pb | 4.8600 | 18.062 | 0.5932 | 18.656 | 0.0873 | 0.1547 |
| Sn-0.012 wt% Pb | 4.6391 | 17.241 | 0.5932 | 17.835 | 0.0778 | 0.1547 |

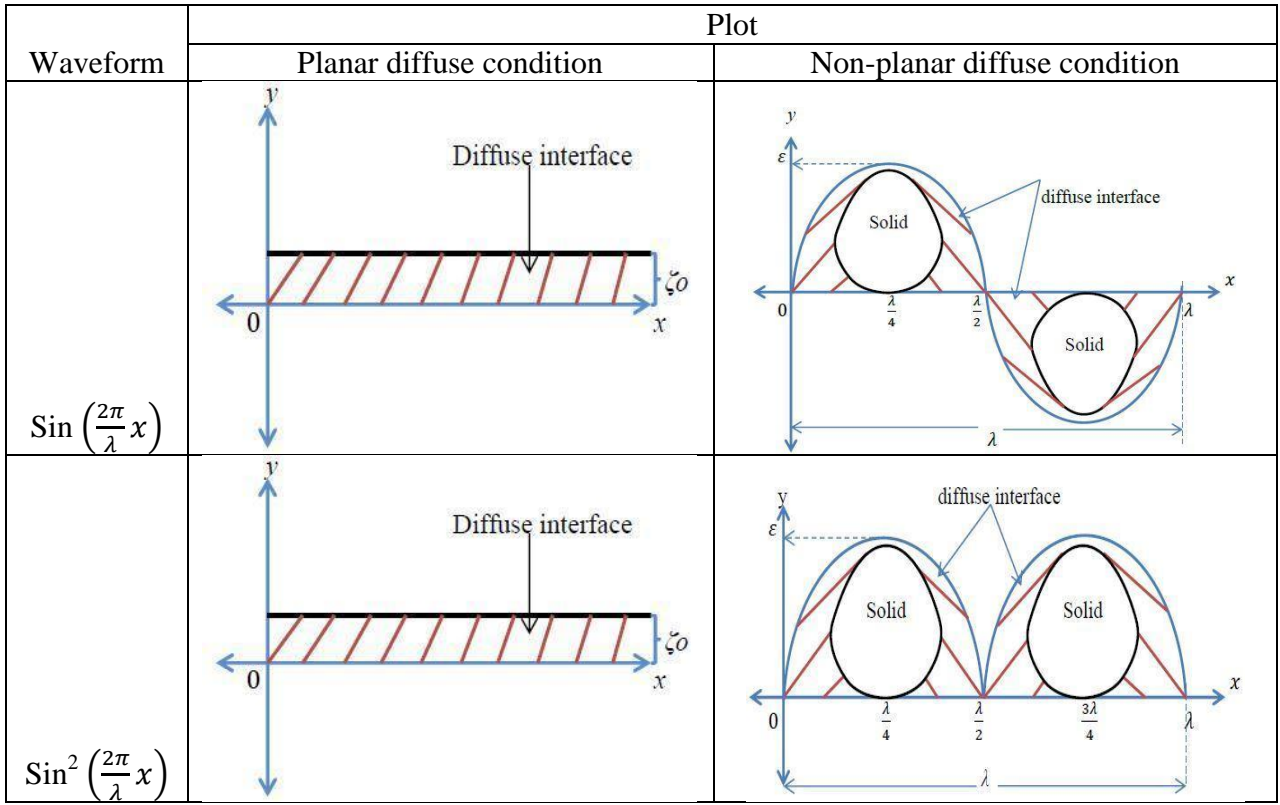

Figure 1. Shows a schematic of plane-front diffuse and non-planar diffuse interfaces for two typical waveforms. The hatched area represents the diffuse interface. The non-planar shapes reach a minimum value for $\zeta$ at the tip of the interface growing into the liquid and reach a maximum as the temperature approaches the solidus temperature, $T_s$. The extent of diffuseness increases towards the root of the shape.

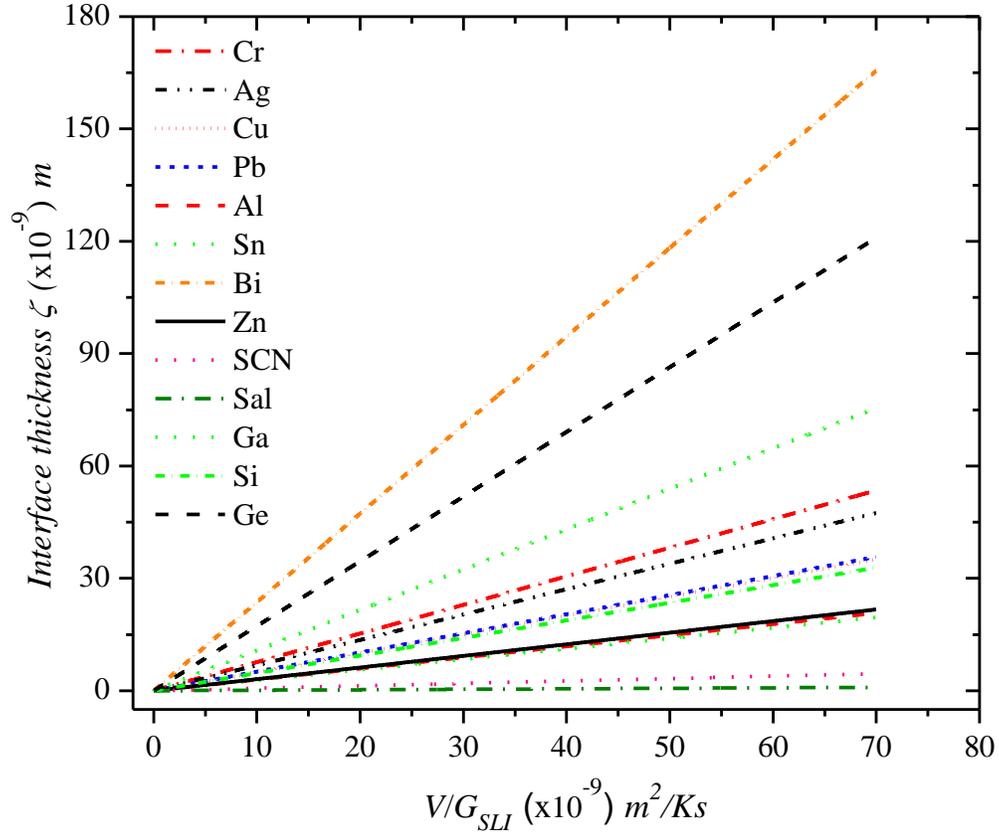

Figure 2. Model prediction for diffuse interface thickness $\zeta$ (m) against $V/G_{SLI}$ ($m^2K^{-1}s^{-1}$) for pure materials as given by equation (33). The diffuse interface thickness is calculated for a fixed temperature gradient and by changing the velocity. The slope of each line is equal to $\mathbf{1}/\sqrt{\mathbf{M}}$ ($K\ s\ m^{-1}$).

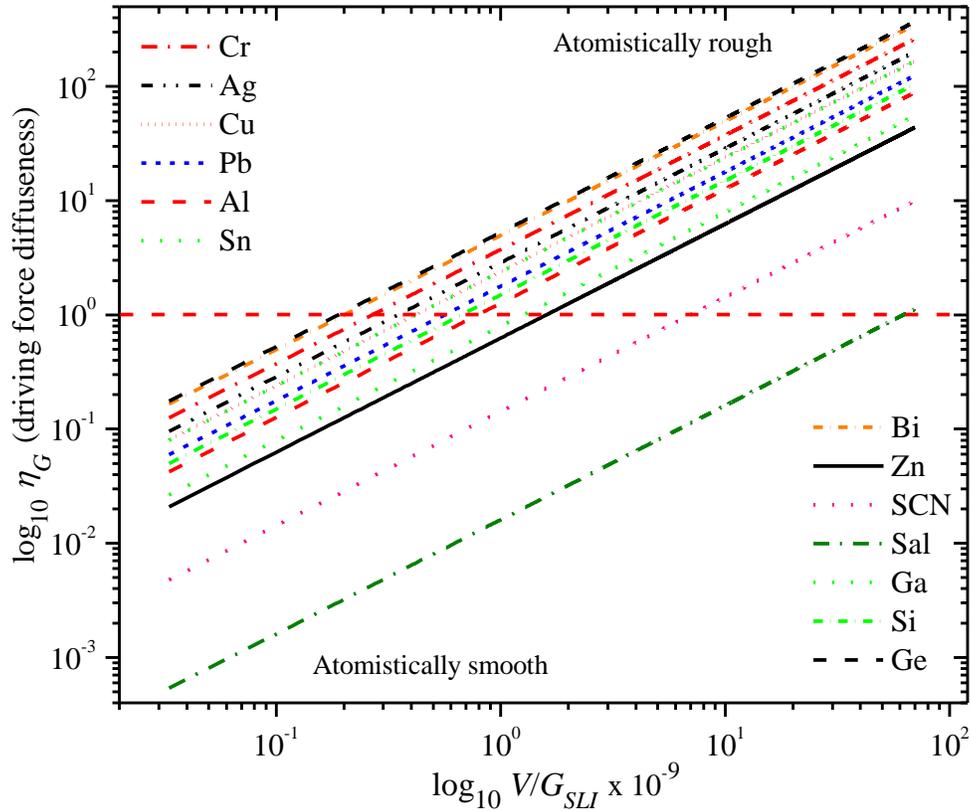

Figure 3. Model prediction for driving force diffuseness $\eta_G$ (dimensionless) as against $V/G_{SLI}$ ($m^2 K^{-1} s^{-1}$) for pure materials showing both atomically smooth and rough interfaces as according to equation (35). The driving force diffuseness is calculated from a fixed temperature gradient and a varied velocity. The dotted red horizontal line indicates one atomic spacing and serves as the criteria between atomically rough and atomically smooth interfaces.

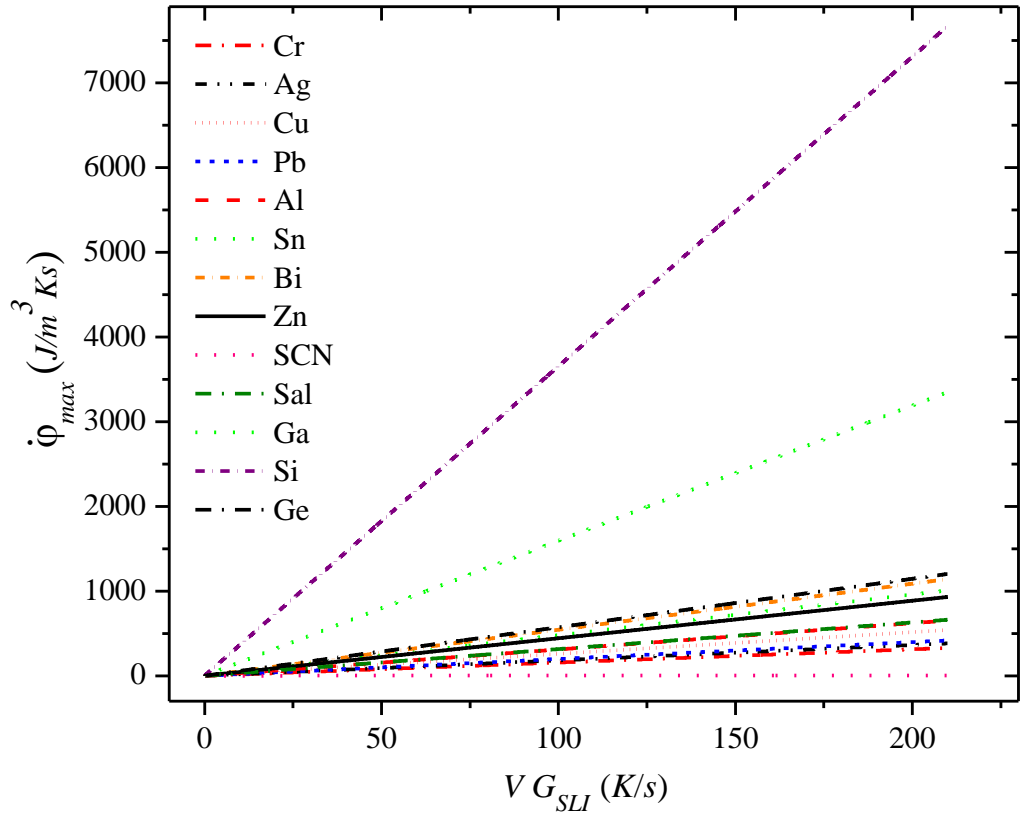

Figure 4. Model prediction for the maximum entropy generation rate density $\dot{\varphi}_{max}$ ($Jm^{-3}K^{-1}s^{-1}$) against $VG_{SLI}$ ($Ks^{-1}$) for pure materials according to equation (36a). The slope of the line is $\Delta S_{sl}/T_m$ ($Jm^{-3}K^{-2}$).

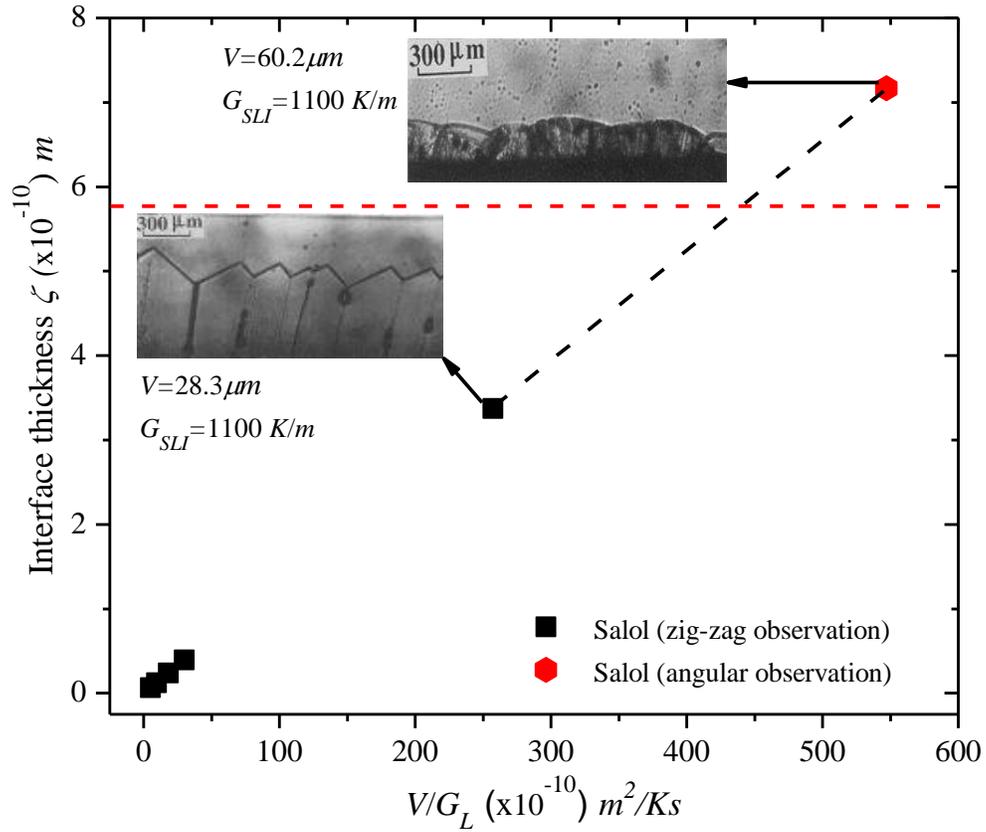

Figure 5. Comparison of calculated $\zeta$ ($m$) against published experimentally measured $V/G_{SLI}$ ($m^2 K^{-1} s^{-1}$) ratio for salol according to equation (33). The value of $\mathbf{1/\sqrt{M}}$ is the slope (0.0131 $K\,s\,m^{-1}$) of the line. The plot shows the transition from facet morphology to non-facet morphology with increasing velocity as shown for dotted black diagonal line. The inserted images [40] are from experiment and show the interface morphologies formed during the transition. The horizontal dotted red line represents the boundary between facet morphology to non-facet morphology.

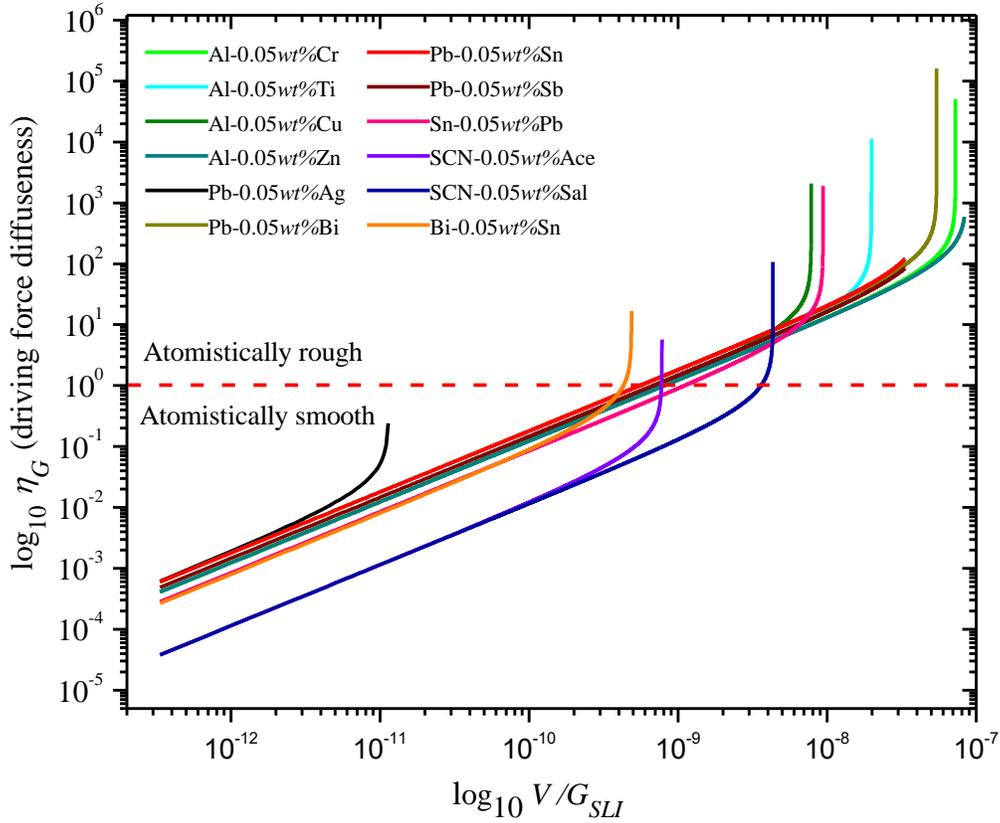

Figure 6. Type Model prediction for the relationship between driving force diffuseness $\eta_G$ and, the ratio of velocity ($V$)/temperature gradient ($G_{SLI}$) for dilute binary materials from equation (44). The dotted horizontal red line indicates the transition line between atomically smooth and atomically rough interface. Materials above the red dashed line have atomically rough interface and materials below have atomically smooth interface. There is no diffuseness at high $V/G_{SLI}$ when $N$ turns zero. The sudden increase in slope at high $V/G_{SLI}$ ratio occurs when $\sqrt{N}$ becomes less than one.

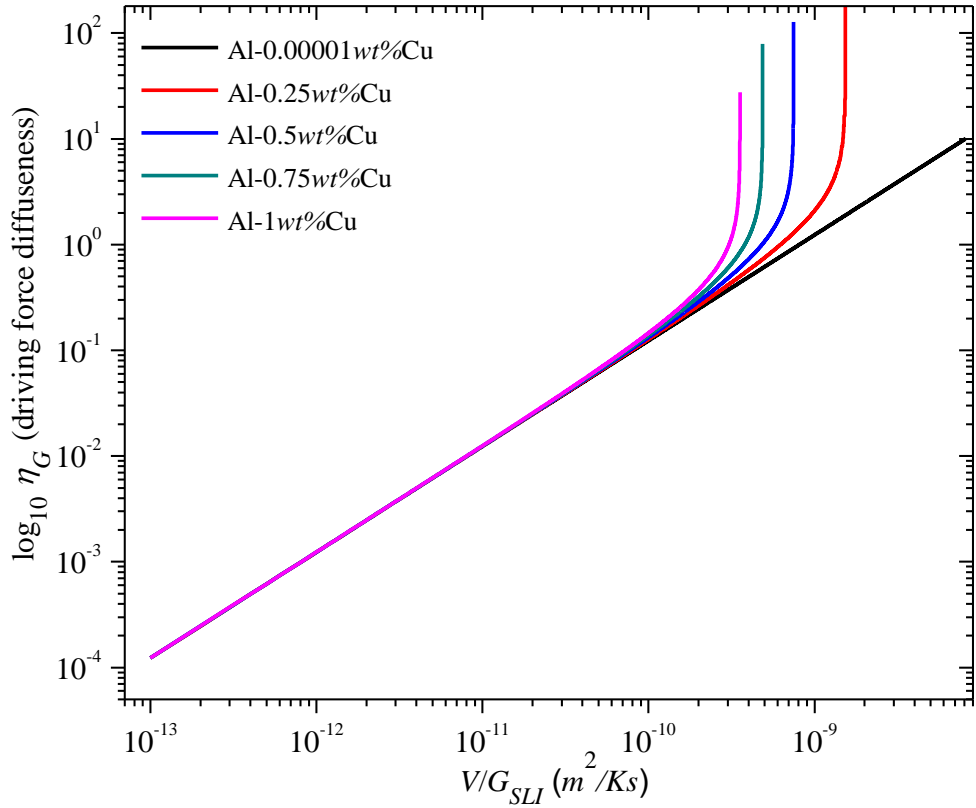

Figure 7. A model prediction for the relationship between *driving force diffuseness* $\eta_G$ and ($V/G_{SLI}$) for Al-Cu at different solute concentrations per equation (44). The plot displays linear forms at low velocities and changes slope at higher velocities. At very low concentration the relationship is pure linear and becomes same as that of a pure material. The linearity is because of the absence of partitioning at the interface. The plot is analogous to figure 5 at fixed solute concentration for different materials.

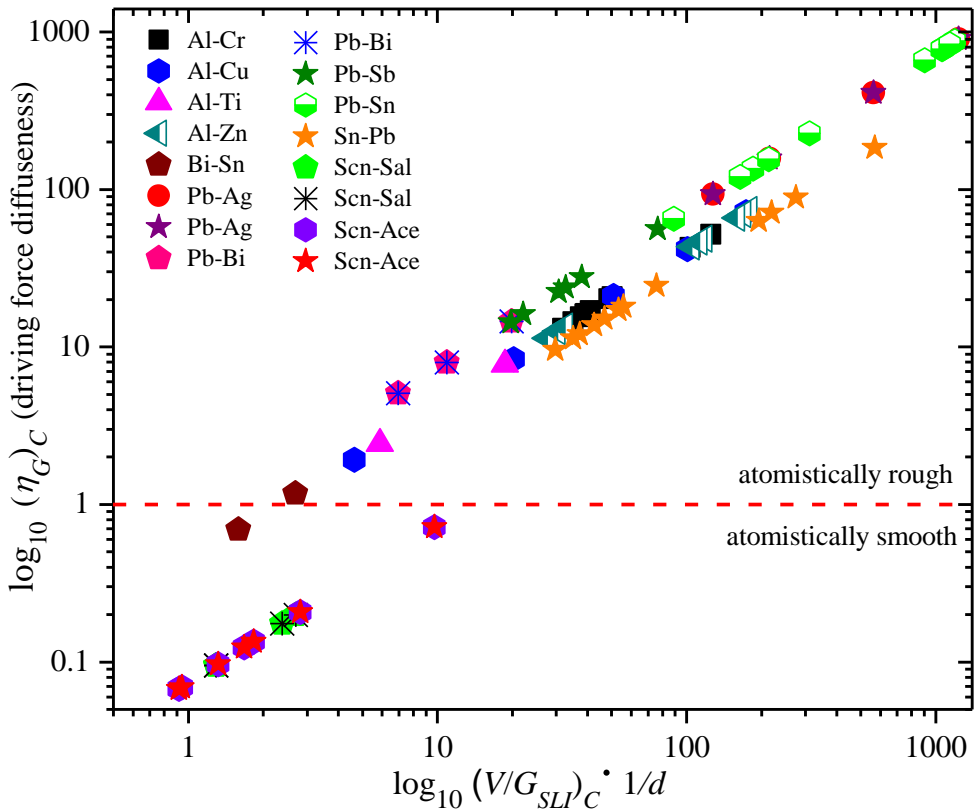

Figure 8. The relationship between driving force diffuseness $(\eta_G)_C$ and, the ratio of velocity $(V_C)$/temperature gradient $(G_{SLI})_C$ by interplanar spacing ($d$). This yields a straight line as per equation (46c) irrespective of material parameters for any growth direction (or crystal plane spacing normal to a growth direction). The plot above shows measured experimental conditions at breakdown in the abscissa and calculated *interface diffuseness* on the ordinate. The horizontal dotted-red line indicates the transition line between atomistically smooth to atomistically rough regimes. Materials above the dotted-red line are atomistically rough and materials below are atomistically smooth. For all metallic materials (in the region above and below the dashed line) only one slope (equal to 0.72995 *K s m⁻¹*) is observed. Also for plastic materials in the region below the dashed line, i.e. the atomistically smooth region, only one slope (equal to 0.07373 *K s m⁻¹*) is observed. In the phase field literature the number of atomic layers in the diffuse region [52], can vary between 2-2750 lattice spacings which are usually an apriory assumption of the interface thickness. From the graph above the diffuse interface are approximately 0.07 to 834 lattice spacings. The calculated driving force diffuseness for this figure is given in table 2.

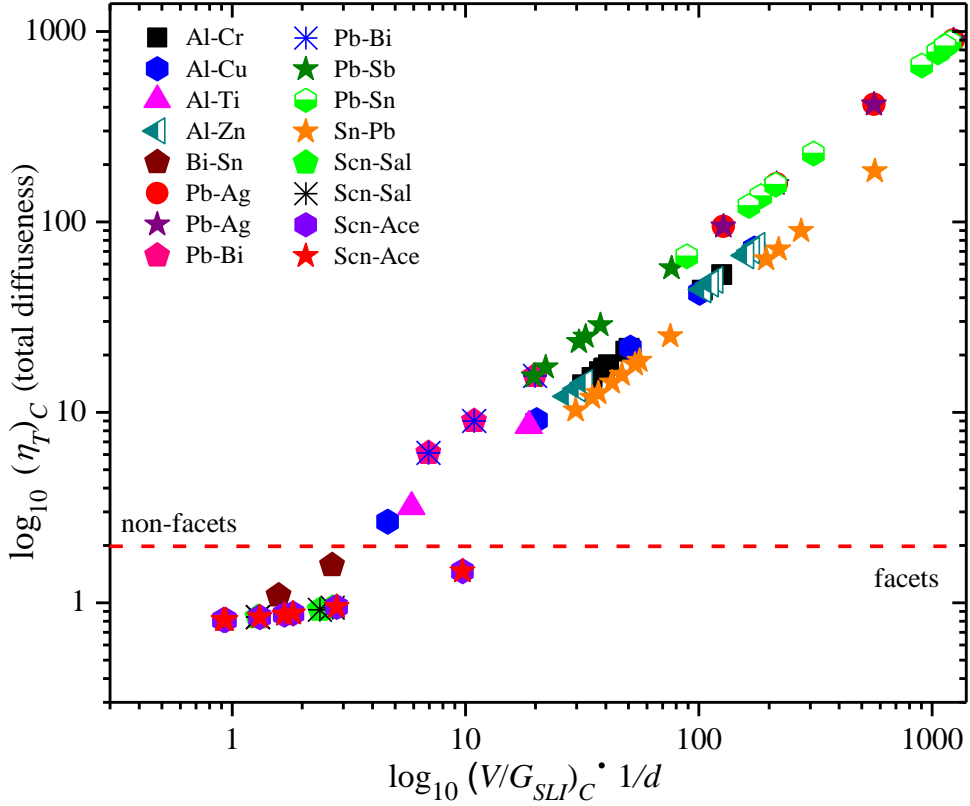

Figure 9. The relationship between total diffuseness and the ratio of velocity $(V)_C$/temperature gradient $(G_{SLI})_C$ should yield a straight line as per equation (48b) irrespective of material parameters for any growth direction (or crystal plane spacing normal to a growth direction). The plot above shows measured experimental conditions at breakdown in the abscissa and calculated interface diffuseness on the ordinate. The *total diffuseness* is the sum of both ($\eta_\alpha + \eta_G$). If the *total diffuseness* is greater than two then there is a possibility of non-facet morphology at breakdown, otherwise it should be facet morphology. The values $V$ and $G_{SLI}$ are experimentally measured numbers at breakdown and $\eta_T$ is calculated from the model. Note that SCN alloys are made nonfacet by the thermal diffuseness at the melting temperature which makes SCN material transformation always appear non-faceted for optical level measurements. Experimentally, the materials shown below the dashed line ($\log_{10} \eta_T$=2) are recorded to be macroscopically faceted. For facet materials zone the different slopes may represent different mechanisms for growth, however this is left to a future study. The calculated total diffuseness for each binary material for this figure is given in table 2.

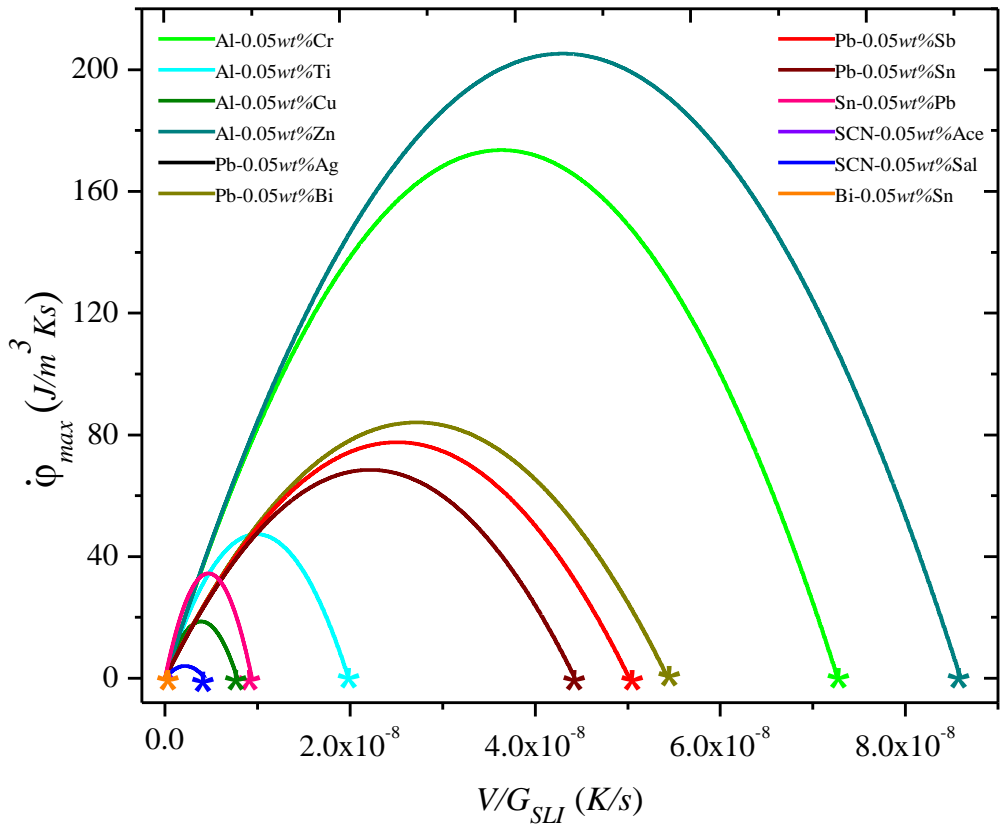

Figure 10. A graph showing model prediction of calculated maximum entropy generation rate density $\boldsymbol{\varphi_{max}}$ ($Jm^{-3}K^{-1}s^{-1}$) against ($V/G_{SLI}$) as per equation (31) for binary materials. At the peak of the curve $\boldsymbol{M}$ is always equal to $2\boldsymbol{B}$ and $\boldsymbol{M}/\sqrt{\boldsymbol{N}}$ is equal to a constant. The star symbol at the end of the curves represents the point where the *diffuse interface* is zero.

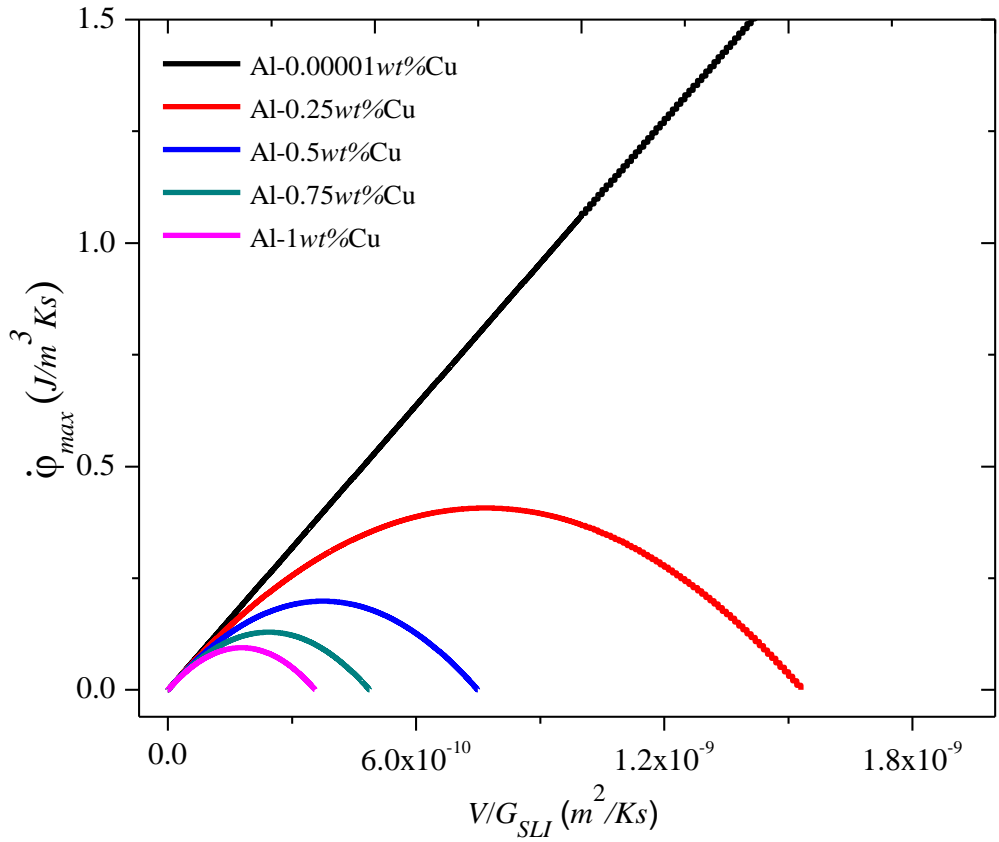

Figure 11. A graph showing the model prediction of calculated $\dot{\boldsymbol{\varphi}}_{\boldsymbol{max}}$ ($Jm^{-3}K^{-1}s^{-1}$) against the ($V/G_{SLI}$) as per equation (31) for Al-Cu at different solute concentrations. The $\dot{\boldsymbol{\varphi}}_{\boldsymbol{max}}$ increases with decreasing solute concentration. At very low solute concentration the binary material behaves like a pure material and $\dot{\boldsymbol{\varphi}}_{\boldsymbol{max}}$ increases indefinitely with $V/G_{SLI}$ ratio as a result of the partition coefficient approaching one.

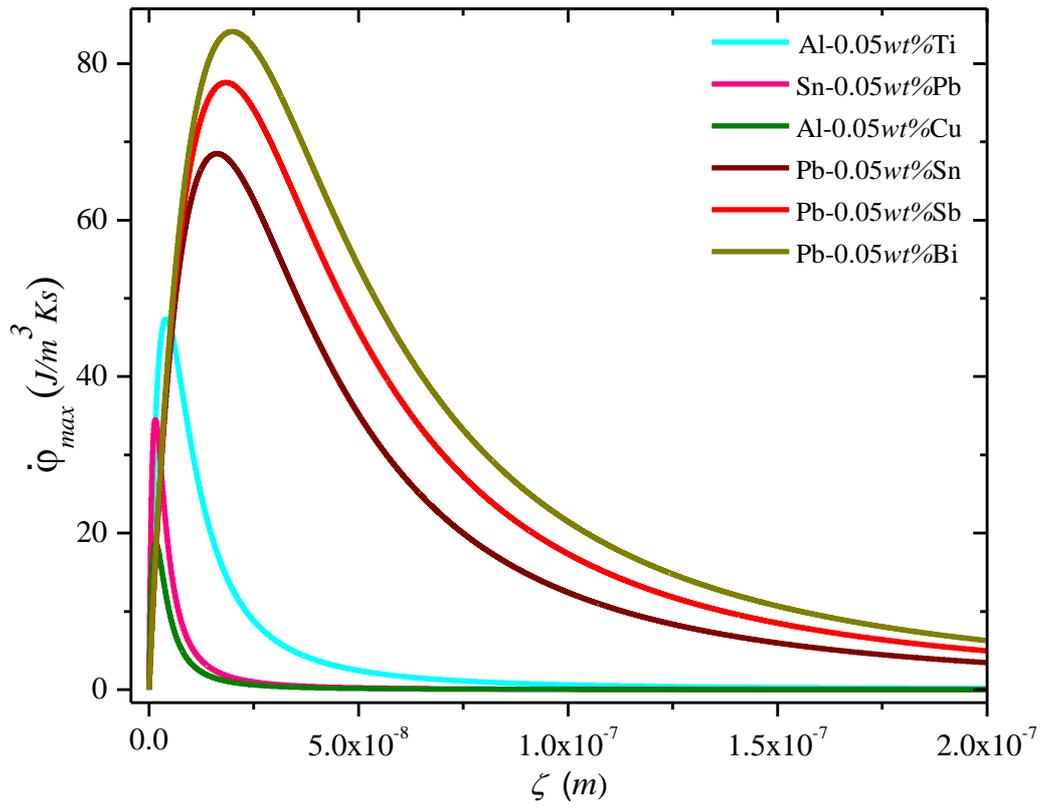

Figure 12. Type Model prediction showing an asymmetric bell shape for a plot of calculated maximum entropy generation rate density $\dot{\boldsymbol{\varphi}}_{max}$ ($Jm^{-3}K^{-1}s^{-1}$) against the *diffuse interface thickness* at a constant solute concentration for different binary materials as per equation (31). As $\dot{\boldsymbol{\varphi}}_{max}$ reaches its highest value at the peak of the curve is when $\boldsymbol{M}$ becomes twice $\boldsymbol{B}$.

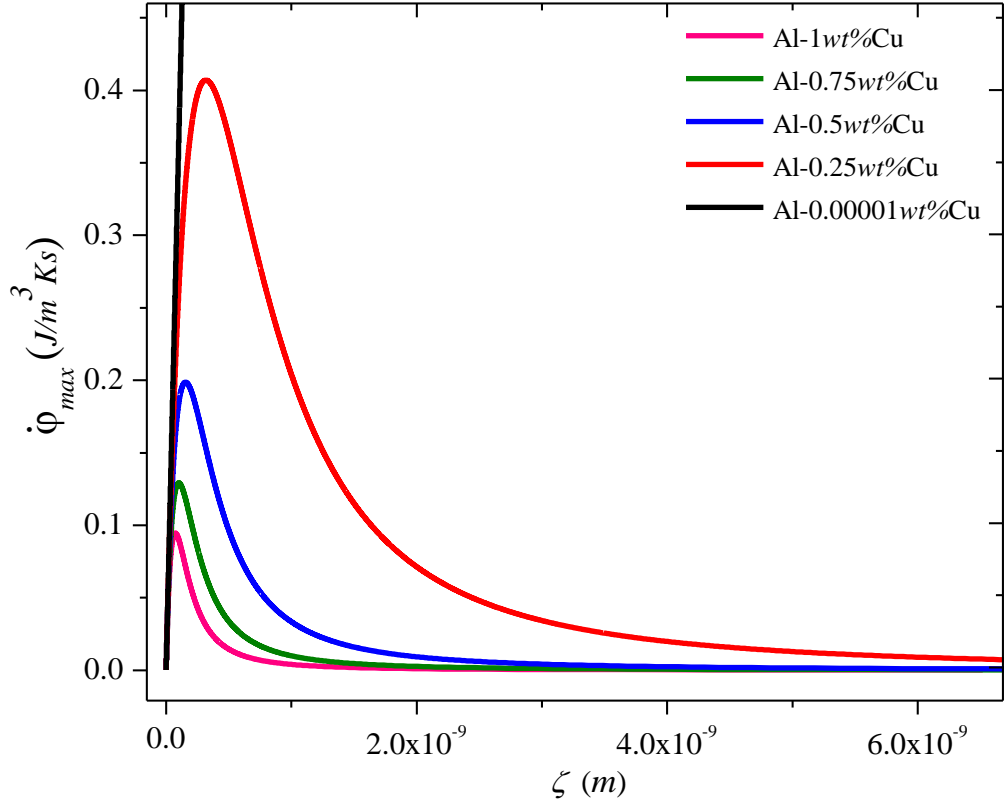

Figure 13. Model prediction showing a plot of calculated maximum entropy generation rate density $\dot{\varphi}_{max}$ ($Jm^{-3}K^{-1}s^{-1}$) against the *diffuse interface thickness* for Al-Cu at different solute concentrations as per equation (31). The plot assumes an asymmetric bell shape and $\dot{\varphi}_{max}$ increases indefinitely with increasing velocity at very dilute solute concentration. The plot is analogous to figure 12 for different binary materials at constant solute concentration.